\documentclass[aps,prd,reprint,twocolumn,superscriptaddress,showpacs]{revtex4-2}

\usepackage{graphicx}
\usepackage{mathrsfs}
\usepackage{bm}
\usepackage{amsmath}
\usepackage{dcolumn}
\usepackage{epstopdf}
\usepackage{dsfont}
\usepackage{amssymb}
\usepackage{tabularx}
\usepackage{array}
\usepackage{float}
\usepackage{color}
\usepackage{epstopdf}
\usepackage{mathrsfs}
\usepackage[colorlinks, linkcolor=blue,anchorcolor=blue,citecolor=blue,urlcolor=blue]{hyperref}
\usepackage{multirow}

\begin{document}

\title{Real Chern Insulators in Two-Dimensional Altermagnetic Fe$_2$S$_2$O and Fe$_2$Se$_2$O}

\author{Yong-Kun Wang}
\affiliation{School of Physics, Northwest University, Xi'an 710127, China}
\affiliation{Shaanxi Key Laboratory for Theoretical Physics Frontiers, Xi'an 710127, China}

\author{Shifeng Qian}
\affiliation{Anhui Province Key Laboratory for Control and Applications of Optoelectronic Information Materials, Department of Physics, Anhui Normal University, Wuhu, Anhui 241000, China}

\author{An-Dong Fan}
\affiliation{School of Physics, Northwest University, Xi'an 710127, China}
\affiliation{Shaanxi Key Laboratory for Theoretical Physics Frontiers, Xi'an 710127, China}

\author{Si Li}
\email{sili@nwu.edu.cn}
\affiliation{School of Physics, Northwest University, Xi'an 710127, China}
\affiliation{Shaanxi Key Laboratory for Theoretical Physics Frontiers, Xi'an 710127, China}

\begin{abstract}
Altermagnets, recently identified as a third class of collinear magnetic materials, have attracted significant attention in condensed matter physics. Despite this growing interest, the realization of real Chern insulators in intrinsic altermagnetic systems has rarely been reported.
In this work, based on first-principles calculations and theoretical analysis, we identify monolayer Fe$_2$S$_2$O and Fe$_2$Se$_2$O as a novel class of two-dimensional altermagnetic real Chern insulators. We demonstrate that these materials possess altermagnetic ground states and host a nontrivial mirror real Chern number, leading to the emergence of symmetry-protected zero-dimensional corner states. Notably, these corner modes are spin-polarized, giving rise to a unique spin–corner coupling effect.
We further show that the real Chern insulating phases and their associated corner states remain robust against spin–orbit coupling, as well as under both uniaxial and biaxial strain. Additionally, these materials exhibit pronounced linear dichroism and strong optical absorption. Our findings uncover the novel topological character of Fe$_2$S$_2$O and Fe$_2$Se$_2$O, establishing them as promising platforms for exploring real Chern insulators in altermagnetic systems.
\end{abstract}

\maketitle
\section{Introduction}
Materials exhibiting nontrivial band topology have become a central focus of research in condensed matter physics~\cite{hasan2010colloquium,qi2011topological,bansil2016colloquium,armitage2018weyl}. The majority of topological states are characterized based on the occupied band eigenstates, which are complex functions. However, under specific symmetry constraints—such as the presence of spacetime inversion symmetry ($\mathcal{PT}$) and the absence of spin–orbit coupling (SOC)—the band eigenstates are constrained to be real~\cite{zhao2016unified,zhao2017pt,ahn2018band,ahn2019stiefel}. In this case, a class of real topological phases can emerge, including real Chern insulators and real semimetals~\cite{zhao2017pt,sheng2019two,lee2020two,chen2022second}. The real Chern insulator is a two-dimensional (2D) topological state characterized by a real Chern number $\nu_R$ (also known as the second Stiefel–Whitney number)~\cite{zhao2017pt,ahn2019stiefel,bouhon2020geometric}, and it typically features symmetry-protected zero-energy modes localized at the corners. To date, several candidate materials for real Chern insulators have been proposed, including graphdiyne~\cite{sheng2019two,lee2020two}, graphyne~\cite{chen2021graphyne}, liganded Xenes~\cite{qian2021second,pan2022two}, monolayer Ta$_2$M$_3$Te$_5$ (M = Ni, Pd)~\cite{guo2022quadrupole}, twisted bilayer graphene~\cite{park2019higher,ahn2019failure}, monolayer Ti$_2$SiCO$_2$~\cite{han2024cornertronics}, and the organic framework material Co$_3$(HITP)$_2$~\cite{zhang2023magnetic}, which have sparked significant interest in this topological phase.

Most recently, the discovery of altermagnetism has established a third fundamental type of magnetic order, extending beyond the conventional ferromagnetic/antiferromagnetic dichotomy~\cite{Naka19NC,Ahn19PRB,Hayami20PRB,yuanLD20PRB,Smejkal2020,Mazin21PNAS,vsmejkal2022beyond,vsmejkal2022emerging,bai2024altermagnetism,fender2025altermagnetism}. Altermagnetic materials exhibit features of both ferromagnets and antiferromagnets: they possess spin-split band structures and macroscopic effects that break time-reversal symmetry, yet retain an antiparallel magnetic configuration with zero net magnetization. A defining characteristic of altermagnetism is that, in real space, the opposite spin sublattices are related by rotational or mirror symmetries, rather than by inversion or translational symmetry~\cite{vsmejkal2022beyond,vsmejkal2022emerging,liu2022spin,xiao2024spin,chen2024enumeration,jiang2024enumeration}. Altermagnetic materials exhibit a variety of unique physical properties, such as the anomalous Hall effect~\cite{Smejkal2020,feng2022anomalous}, unconventional spin current generation~\cite{CJWu07PRB,ma2021multifunctional,gonzalez2021efficient,Bose22NE}, giant tunnelling magnetoresistance~\cite{shao2021spin,vsmejkal2022giant}, spin Seebeck and spin Nernst effects~\cite{cui2023efficient}, unconventional Andreev reflection\cite{Papaj23PRB,sun2023andreev}, finite-momentum Cooper pairing~\cite{Zhang2024Finite,hong2024unconventional,sim2025pair,PhysRevB.110.L060508}, topological superconductivity~\cite{li2023majorana,PhysRevLett.133.106601,PhysRevB.109.L201109,zhu2023topological}, ferroelectric and antiferroelectric~\cite{gu2025ferroelectric,duan2025antiferroelectric,vsmejkal2024altermagnetic,urru2025g}. To date, numerous altermagnetic materials have been identified, including RuO$_2$~\cite{berlijn2017itinerant,zhu2019anomalous,fedchenko2024observation}, MnTe~\cite{gonzalez2023spontaneous,krempasky2024altermagnetic}, MnTe$_2$~\cite{zhu2024observation}, CrSb~\cite{li2024topological,lu2025signature,ding2024large,zhou2025manipulation,yang2025three}, Rb$_{1-\delta}$V$_2$Te$_2$O~\cite{zhang2025crystal}, and KV$_2$Se$_2$O~\cite{jiang2025metallic}.
Altermagnetism offers a fertile ground for realizing magnetic topological quantum states. Nevertheless, real Chern insulators in altermagnetic systems remain rarely reported, with only a few known examples, including monolayer Mg(CoN)$_2$~\cite{han2025real}, MnS$_2$~\cite{wang2025pentagonal}, and iron oxyhalides~\cite{wang2025two}. This scarcity underscores the urgent need to discover and explore additional real Chern insulators in altermagnetic materials.

In this work, through first-principles calculations and theoretical analysis, we reveal monolayer Fe$_2$S$_2$O and Fe$_2$Se$_2$O as a new class of a two-dimensional (2D) altermagnetic real Chern insulators. We show that these materials possess altermagnetic ground states with spin-split band structures in the absence of SOC. Moreover, we find that each spin channel carries a nontrivial mirror real Chern number, $\nu_R^\uparrow = \nu_R^\downarrow = 1$, which is enabled by the preserved $\mathcal{M}_z$ and $\mathcal{P}$ symmetries, leading to the emergence of topologically protected 0D corner states. Notably, these corner modes are spin-polarized, giving rise to a unique spin–corner coupling effect. We demonstrate that the real Chern insulating phases and their associated corner states in monolayer Fe$_2$S$_2$O and Fe$_2$Se$_2$O remain robust against spin–orbit coupling (SOC) as well as under both uniaxial and biaxial strain. Moreover, we find that these materials display pronounced linear dichroism and strong optical absorption. Our findings uncover the intrinsic topological nature of monolayer Fe$_2$S$_2$O and Fe$_2$Se$_2$O, establishing them as ideal material platforms for investigating real Chern topological phases in altermagnetic materials.

\section{Computation Methods}
The first-principles calculations based on density functional theory (DFT) were performed using the projector augmented-wave method, as implemented in the Vienna \emph{ab initio} simulation package (VASP)~\cite{kresse1994,kresse1996,blochl1994projector}. The exchange–correlation potential was treated using the generalized gradient approximation (GGA) with the Perdew–Burke–Ernzerhof (PBE) functional~\cite{PBE}. A plane-wave energy cutoff of 500 eV was used, and the Brillouin zone was sampled with a $\Gamma$-centered Monkhorst–Pack (MP) $17 \times 17 \times 1$ $k$-point mesh. The convergence thresholds were set to $10^{-8}$ eV for total energy and $10^{-3}$ eV/Å for ionic forces. To eliminate spurious interactions between periodic images, a vacuum spacing of 20 Å was applied along the out-of-plane direction.
The on-site Coulomb interaction for the Fe 3$d$ orbitals was accounted for using the DFT+$U$ method~\cite{Anisimov1991,dudarev1998}. The effective $U$ value was taken to be 4 eV for Fe~\cite{jain2011formation}. The test of other $U$ values is presented in Supplemental Material~\cite{SM}). Phonon spectra were computed using a $4 \times 4 \times 1$ supercell within the framework of density functional perturbation theory (DFPT), as implemented in the PHONOPY code~\cite{togo2015first}. To analyze the topological properties, we constructed tight-binding Hamiltonians using maximally localized Wannier functions via the Wannier90 package~\cite{pizzi2020wannier90}, from which the nanodisk spectrum was subsequently calculated.

\section{Results and Discussion}
\subsection{Crystal structure and magnetism}
\begin{figure}[htb]
		\centering
	\includegraphics[width=8.6cm]{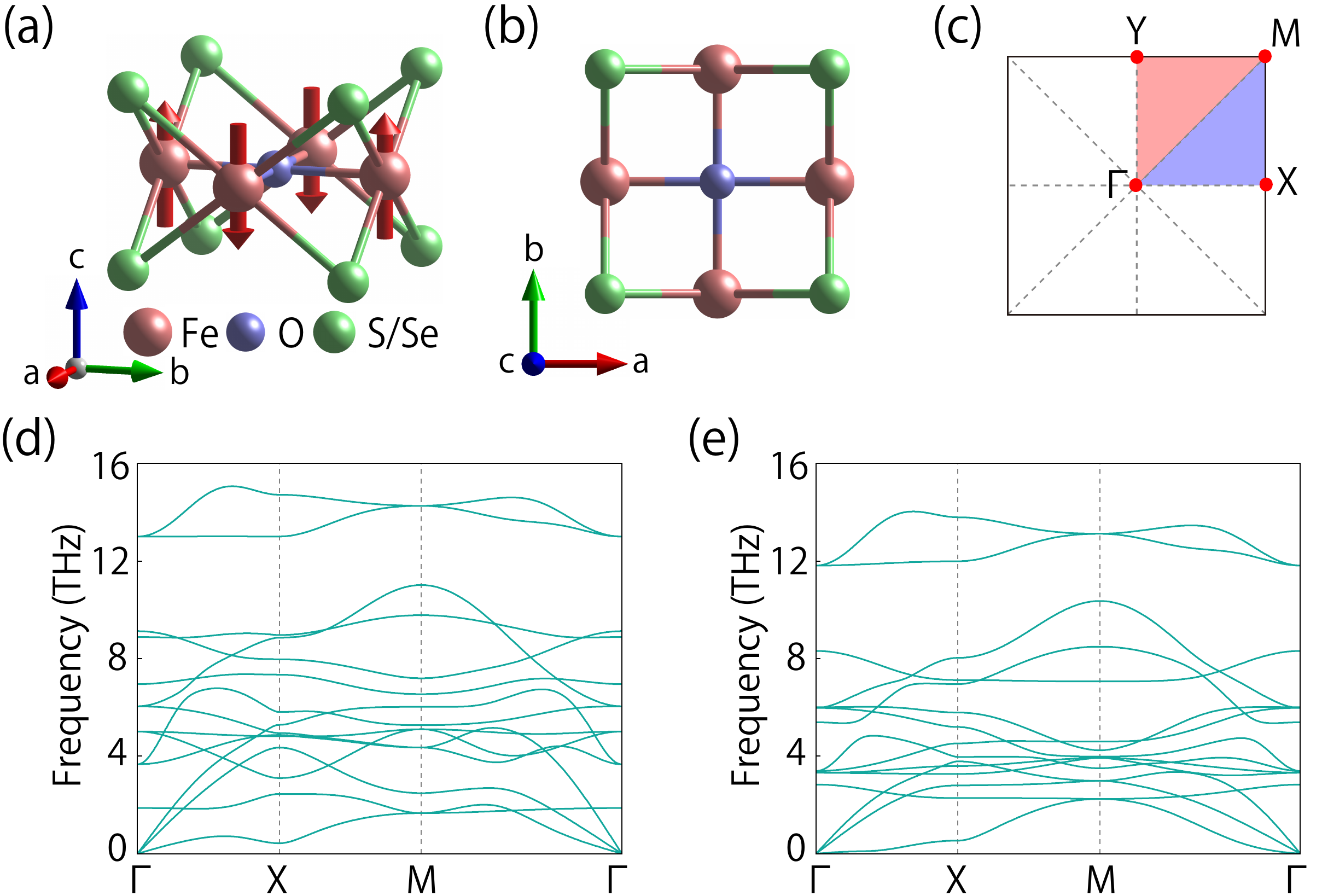}
	\caption{(a) Side view and (b) top view of the crystal structure of monolayer Fe$_2$S$_2$O and Fe$_2$Se$_2$O. The altermagnetic configuration is illustrated in (a). (c) Brillouin zone of the monolayer structure, with high-symmetry points indicated.  
	(d) and (e) Calculated phonon spectra of Fe$_2$S$_2$O and Fe$_2$Se$_2$O, respectively.		
	\label{fig1}}
\end{figure}
Monolayer Fe$_2$S$_2$O and Fe$_2$Se$_2$O feature three-layer atomic structures, where a central layer composed of Fe and O atoms is sandwiched between two outer atomic planes of S and Se, respectively. Both materials crystallize in a tetragonal lattice belonging to the space group P4/mmm (No. 123), as shown in Figs.~\ref{fig1}(a) and~\ref{fig1}(b). The optimized in-plane lattice constants are $a = b = 4.00$ Å for Fe$_2$S$_2$O and $a = b = 4.05$ Å for Fe$_2$Se$_2$O. The dynamical stability of these structures is confirmed by phonon spectrum calculations. As shown in Figs.~\ref{fig1}(d) and~\ref{fig1}(e), the absence of imaginary frequencies indicates that both monolayers are dynamically stable and can exist as freestanding 2D materials. It is worth emphasizing that our proposed monolayer Fe$_2$S$_2$O and Fe$_2$Se$_2$O adopt the same crystal structure as V$_2$Se$_2$O and V$_2$Te$_2$O~\cite{lin2018structure,ablimit2018v2te2o}, which have already been synthesized and have attracted considerable attention in recent studies. This structural similarity suggests that monolayer Fe$_2$S$_2$O and Fe$_2$Se$_2$O could also be promising candidates for experimental synthesis.

Given that monolayer Fe$_2$S$_2$O and Fe$_2$Se$_2$O contain the $3d$ transition metal element Fe, which typically exhibits magnetic behavior, we begin by determining their magnetic ground state. This is achieved by comparing the total energies of several representative magnetic configurations, including ferromagnetic (FM), altermagnetic (AM), and two antiferromagnetic (AFM) states (see the Supplemental Material~\cite{SM}). Our calculations indicate that both Fe$_2$S$_2$O and Fe$_2$Se$_2$O energetically favor the altermagnetic configuration [as illustrated in Fig.~\ref{fig1}(a)]. In this altermagnetic ground state, the magnetic moments are predominantly localized on the Fe atoms, with a magnitude of approximately \(4~\mu_B\) per Fe site. The two spin-opposed sublattices are connected through the combined $\mathcal{C}_{4z}\mathcal{T}$ symmetry. 
In the absence of SOC, monolayer Fe$_2$S$_2$O and Fe$_2$Se$_2$O belongs to spin space group (SSG) No. 47.123.1.1. 

\subsection{Bulk band structure}
\begin{figure*}[htb]
		\centering
	\includegraphics[width=16cm]{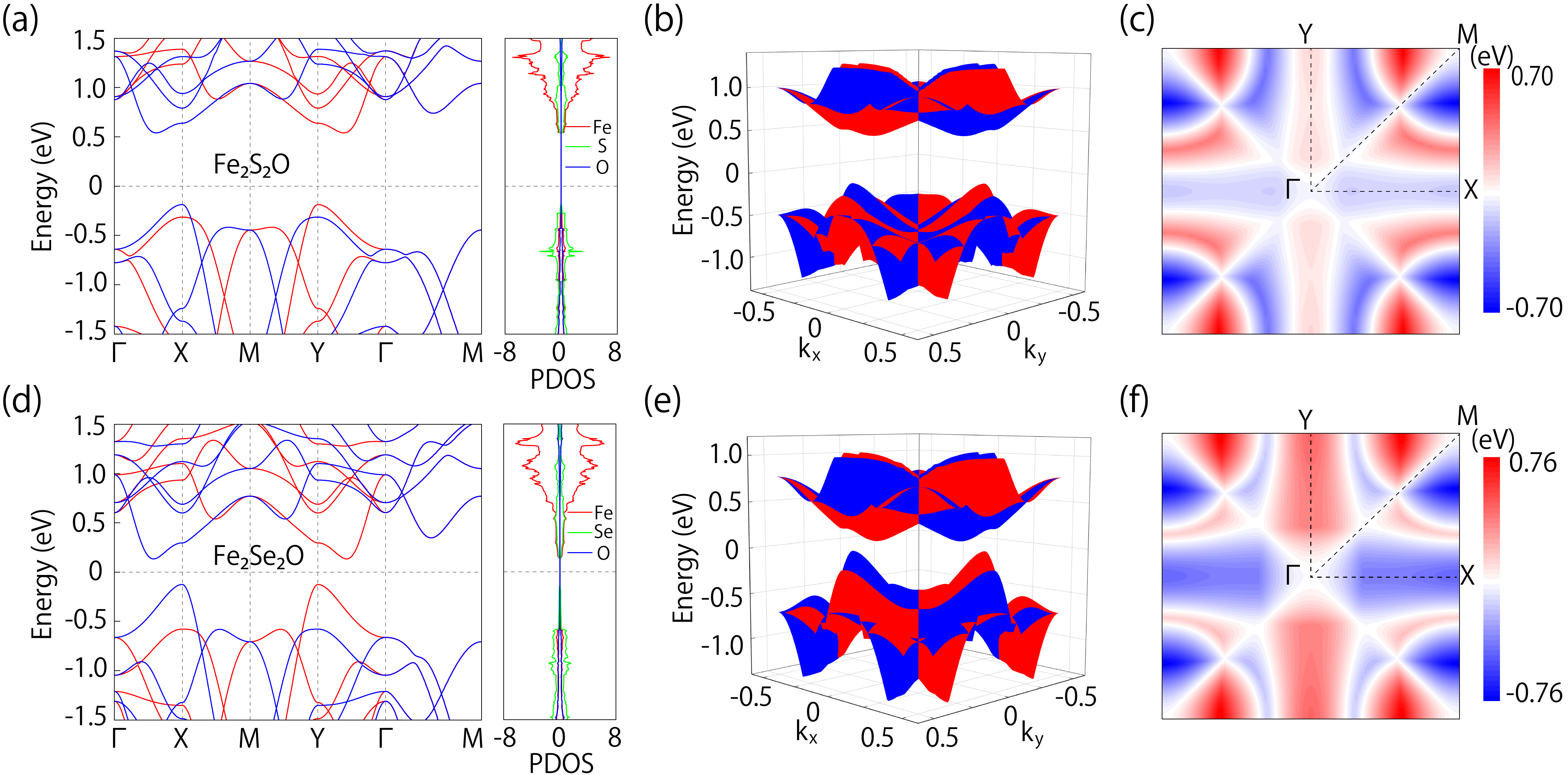}
	\caption{(a) Band structure and PDOS; (b) two-dimensional band structures of the two lowest conduction bands and the two highest valence bands; and (c) spin splitting of the two highest valence bands for monolayer Fe$_2$S$_2$O. (d)–(f) Corresponding results for monolayer Fe$_2$Se$_2$O. In (a), (b), (d) and (e), red and blue represent spin-up and spin-down bands, respectively. The SOC is not included.
	\label{fig2}}
\end{figure*}
In this section, we investigate the electronic band structures of monolayer Fe$_2$S$_2$O and Fe$_2$Se$_2$O in their altermagnetic ground state. Since SOC has a negligible effect on the band structures of these materials, we mainly focus on the band structures without SOC. The band structures with SOC included are provided in section~\ref{soc}. The band structures and projected density of states (PDOS) of monolayer Fe$_2$S$_2$O and Fe$_2$Se$_2$O without SOC are shown in Figs.~\ref{fig2}(a) and~\ref{fig2}(d). As shown in Figs.~\ref{fig2}(a) and~\ref{fig2}(d), monolayer Fe$_2$S$_2$O and Fe$_2$Se$_2$O are indirect bandgap semiconductors, with the valence band maximum (VBM) located at the X and Y points and the conduction band minimum (CBM) along the $\Gamma$–X and $\Gamma$–Y paths. The band gaps without SOC are 0.72 eV for Fe$_2$S$_2$O and 0.26 eV for Fe$_2$Se$_2$O. The PDOS reveals that the low-energy bands are predominantly contributed by the S/Se atoms. Moreover, their band structures exhibit momentum-dependent spin splitting in the absence of SOC, which is a hallmark of altermagnetism. 
The 2D band structures of the two lowest conduction bands and two highest valence bands for monolayer Fe$_2$S$_2$O and Fe$_2$Se$_2$O are shown in Figs.~\ref{fig2}(b) and~~\ref{fig2}(e), respectively. It is evident that bands with opposite spins exhibit $\mathcal{C}_{4z}\mathcal{T}$ symmetry. The spin splitting of the two highest valence bands for monolayer Fe$_2$S$_2$O and Fe$_2$Se$_2$O are shown in Figs.~\ref{fig2}(c) and~~\ref{fig2}(f), respectively. From Figs.~\ref{fig2}(c) and~\ref{fig2}(f), one can observe that both systems exhibit significant spin splitting, reaching up to 0.7 eV.

 In the absence of SOC, the system preserves the symmetries $\mathcal{C}_{4z}\mathcal{T}$, $\mathcal{P}$, and $\mathcal{M}_z$. In 2D systems, real Chern insulators with inversion symmetry $\mathcal{P}$ can generally be characterized by the real Chern number $\nu_R$, also known as the second Stiefel–Whitney number~\cite{zhao2017pt,ahn2019stiefel}. It is computed as:
\begin{eqnarray}
	(-1)^{\nu_R} = \prod_i (-1)^{\lfloor n_-^{(\Gamma_i)} / 2 \rfloor}, \label{RCM}
\end{eqnarray}  
where $\lfloor...\rfloor$ is the floor function, and \( n_-^{(\Gamma_i)} \) is the number of valence states with negative parity eigenvalues at the time-reversal invariant momentum (TRIM) points $\Gamma_i$. A nontrivial topological phase is indicated by \( \nu_R = 1 \), whereas \( \nu_R = 0 \) corresponds to a trivial phase.  

\begin{table}[htb]
	\centering
	\caption{\label{table1} Parity information at the four TRIM points of monolayer Fe$_2$S$_2$O and Fe$_2$Se$_2$O. The coordinates of these points are given by $\Gamma$ (0, 0), X (0.5, 0), Y (0, 0.5), and M (0.5, 0.5). Here, $n_+$ ($n_-$) denotes the number of occupied bands with positive (negative) parity eigenvalues. The real Chern numbers $\nu_{R}^{\uparrow}$ and $\nu_{R}^{\downarrow}$ are both equal to 0.}
	\begin{tabular}{cccccccccc}
		\hline
		\multicolumn{6}{c}{Spin-Up} & 
		\multicolumn{4}{c}{Spin-Down} \\ 
		\cline{2-5}
		\cline{7-10}
		& $\Gamma$ & X & Y & M & & $\Gamma$ & X & Y & M \\ 
		\hline
		$n_+$ & 10 & 7 & 12 & 5 & & 10 & 12 & 7 & 5 \\
		\hline
		$n_-$ & 7 & 10 & 5 & 12 & & 7 & 5 & 10 & 12 \\
		\hline
		\multicolumn{6}{c}{$\nu_{R}^{\uparrow} = 0$} & 
		\multicolumn{4}{c}{$\nu_{R}^{\downarrow} = 0$} \\	
		\hline
	\end{tabular}	
\end{table}

\begin{table}[htb]
	\centering
	\renewcommand\arraystretch{1.5}
	\caption{\label{table2} Mirror-resolved $\mathcal{P}$ eigenvalues of all occupied bands at $\Gamma$, X, Y, and M points of the monolayer Fe$_2$S$_2$O and Fe$_2$Se$_2$O. Here, $n_+$ ($n_-$) denotes the number of occupied bands with positive (negative) parity eigenvalues. The real Chern number obtained for each spin channel is 	$\nu_R^\uparrow = \nu_R^\downarrow = \nu_R^+ + \nu_R^- = 1$.}
	\begin{tabular}{cccccccccccccccccccccccc}
		\hline
		\multicolumn{12}{c}{Spin-Up} &  
		\multicolumn{12}{c}{Spin-Down} \\
		\hline
		\multicolumn{7}{c}{$M_z = 1$} & 
		\multicolumn{4}{c}{$M_z = -1$}&
		\multicolumn{9}{c}{$M_z = 1$} &
		\multicolumn{4}{c}{$M_z = -1$} \\
		\cline{2-5}
		\cline{8-11}
		\cline{15-18}
		\cline{21-24}
		
		& $\Gamma$ & X & Y & M &&& $\Gamma$ & X & Y & M &&&&$\Gamma$ & X & Y & M &&& $\Gamma$ & X & Y & M \\ 
		\hline
		$n_+$ & 6 & 4 & 7 & 3 &&& 4 & 3 & 5 & 2 &&& $n_+$ & 6 & 7 & 4 & 3 &&&  4 & 5 & 3 & 2  \\
		\hline
		$n_-$ & 4 & 6 & 3 & 7 &&& 3 & 4 & 2 & 5 &&& $n_-$ & 4 & 3 & 6 & 7 &&&  3 & 2 & 4 & 5  \\
		\hline
		\multicolumn{7}{c}{$\nu_R^+ = 1$} & 
		\multicolumn{4}{c}{$\nu_R^- = 0$} &
		\multicolumn{9}{c}{$\nu_R^+ = 1$} & 
		\multicolumn{4}{c}{$\nu_R^- = 0$} \\
		\hline		
	\end{tabular}
\end{table}

When SOC is neglected, the spin-up and spin-down channels in monolayer Fe$_2$S$_2$O and Fe$_2$Se$_2$O decouple but remain related by the combined $\mathcal{C}_{4z}\mathcal{T}$ symmetry. Each channel can thus be treated as an effective spinless system preserving time-reversal symmetry \( \mathcal{T} \). Although the full $\mathcal{C}_{4z}$ symmetry (and hence space group SG 123) is broken in each spin channel, a detailed symmetry analysis reveals that both channels independently preserve SG 47, generated by $\mathcal{C}_{2x}$, $\mathcal{C}_{2z}$, and $\mathcal{P}$. Due to the decoupling of the spin channels, a real Chern number \( \nu_R^\sigma \) (where \( \sigma = \uparrow, \downarrow \)) can be defined for each spin channel individually.  Table~\ref{table1} lists the $\mathcal{P}$ eigenvalues of the valence bands at the four TRIM. Substituting these values into Eq.~(\ref{RCM}) yields \( \nu_R^\uparrow = \nu_R^\downarrow = 0 \), seemingly suggesting a topologically trivial phase. However, this conclusion does not hold for systems with mirror symmetry. In monolayer Fe$_2$S$_2$O and Fe$_2$Se$_2$O, the presence of horizontal mirror symmetry $\mathcal{M}_z$ allows each spin channel to be decomposed into two independent mirror subspaces:  
\begin{eqnarray}
	\mathcal{H}_\sigma = \mathcal{H}_\sigma^+ \oplus \mathcal{H}_\sigma^-,  
\end{eqnarray}  
where \( \mathcal{H}_\sigma^\pm \) corresponds to subspaces with mirror eigenvalues ( $\mathcal{M}_z = \pm 1$). Each mirror subspace retains both $\mathcal{P}$ and \( \mathcal{T} \) symmetries, enabling the evaluation of its real Chern number via Eq.~(\ref{RCM}). As summarized in Table~\ref{table2}, for both spin channels, the $\mathcal{M}_z = + 1$ subspace exhibits a nontrivial real Chern number \( \nu_R^+ = 1 \), while the $\mathcal{M}_z = - 1$ subspace remains trivial with \( \nu_R^- = 0 \). Consequently, the total real Chern number for each spin channel is given by  
\begin{eqnarray}
	\nu_R^\uparrow = \nu_R^\downarrow = \nu_R^+ + \nu_R^- = 1,  
\end{eqnarray}  
confirming that monolayer Fe$_2$S$_2$O and Fe$_2$Se$_2$O are 2D real Chern insulators.

\subsection{Corner states}
\begin{figure}[htb]
	\centering
	\includegraphics[width=8.6cm]{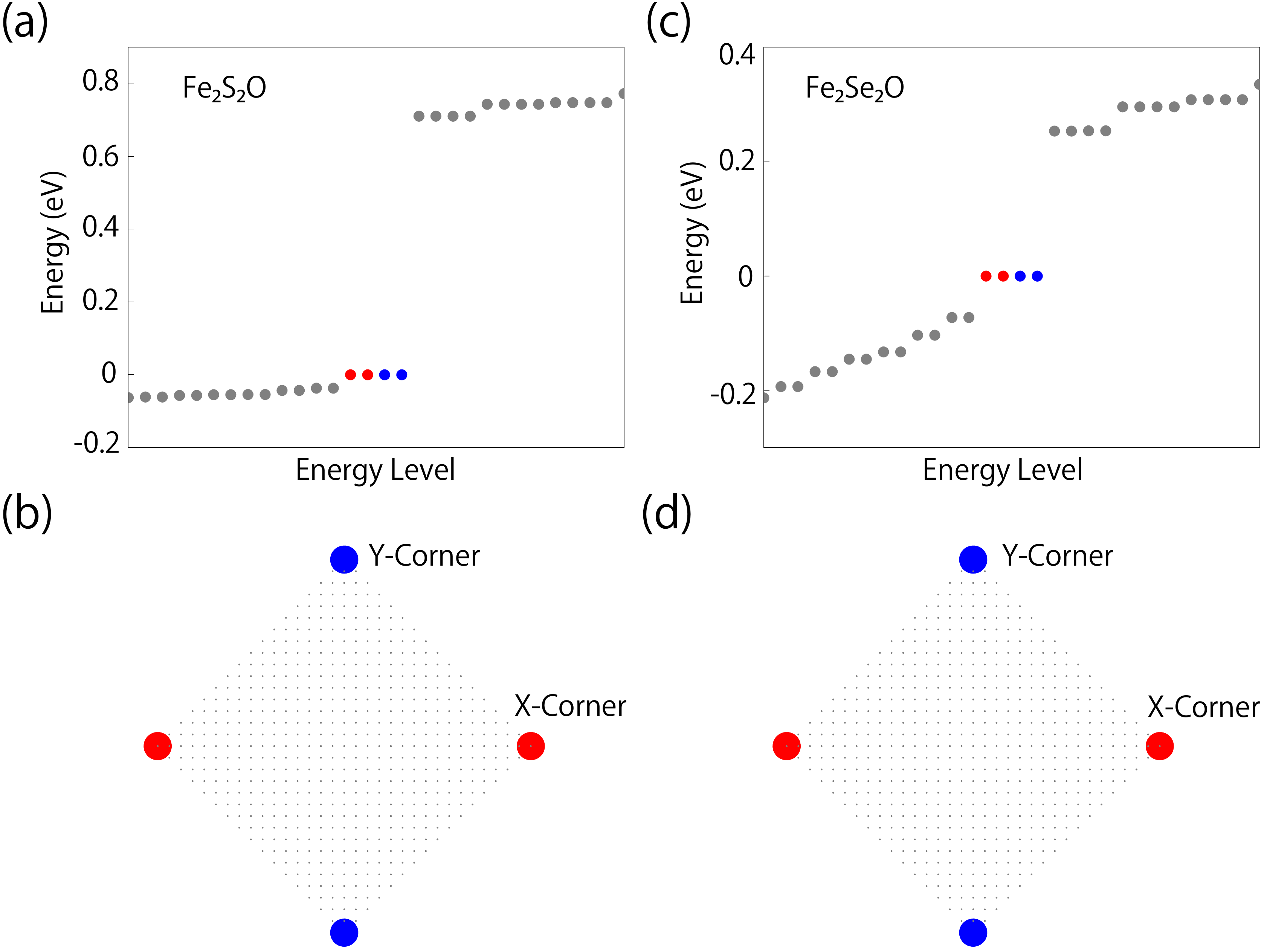}
	\caption{(a) Energy spectrum of the square-shaped Fe$_2$S$_2$O nanodisk shown in (b), with energy levels arranged in ascending order. Four zero-energy states are highlighted in color, where red (blue) dots represent spin-up (spin-down) components. (b) Charge distribution of the four zero-energy states, showing clear localization at the corners. (c) and (d) Corresponding results for the monolayer Fe$_2$Se$_2$O nanodisk.
		\label{fig3}}
\end{figure}

The bulk topology of a real-Chern insulator manifests as zero-energy modes localized at its corners. In any sample geometry that preserves $\mathcal{P}$ symmetry, these corner states necessarily come in $\mathcal{P}$-related pairs, independent of the edge terminations or overall shape. Furthermore, because the spin-up and spin-down sectors are interchanged by the $\mathcal{C}_{4z}\mathcal{T}$ operation, one generically obtains four midgap corner states--two originating from the spin-up component and two from the spin-down component.
To illustrate this, we constructed a square-shaped nanodisk for monolayer Fe$_2$S$_2$O and Fe$_2$Se$_2$O and calculated its energy spectrum and corner states. The results are presented in Fig.~\ref{fig3}. From Fig.~\ref{fig3}, one clearly observes four zero-modes at the Fermi level (each spin has two zero-energy states), isolated within the bulk bandgap. Spatial profiles of these states confirm their corner localization. Intriguingly, the two corners along the $ x $-axis host spin-up states (X corners), whereas the two along the $y$-axis host spin-down states (Y corners). This corner-contrasted spin polarization directly couples real-space position with spin, giving rise to a novel spin-corner coupling effect. The spatial separation of corner states allows, in principle, independent spin control at each corner, potentially leading to distinct experimental signatures.

\subsection{Effects of SOC}{\label{soc}}

\begin{figure}[htb]
	\centering
	\includegraphics[width=8.6cm]{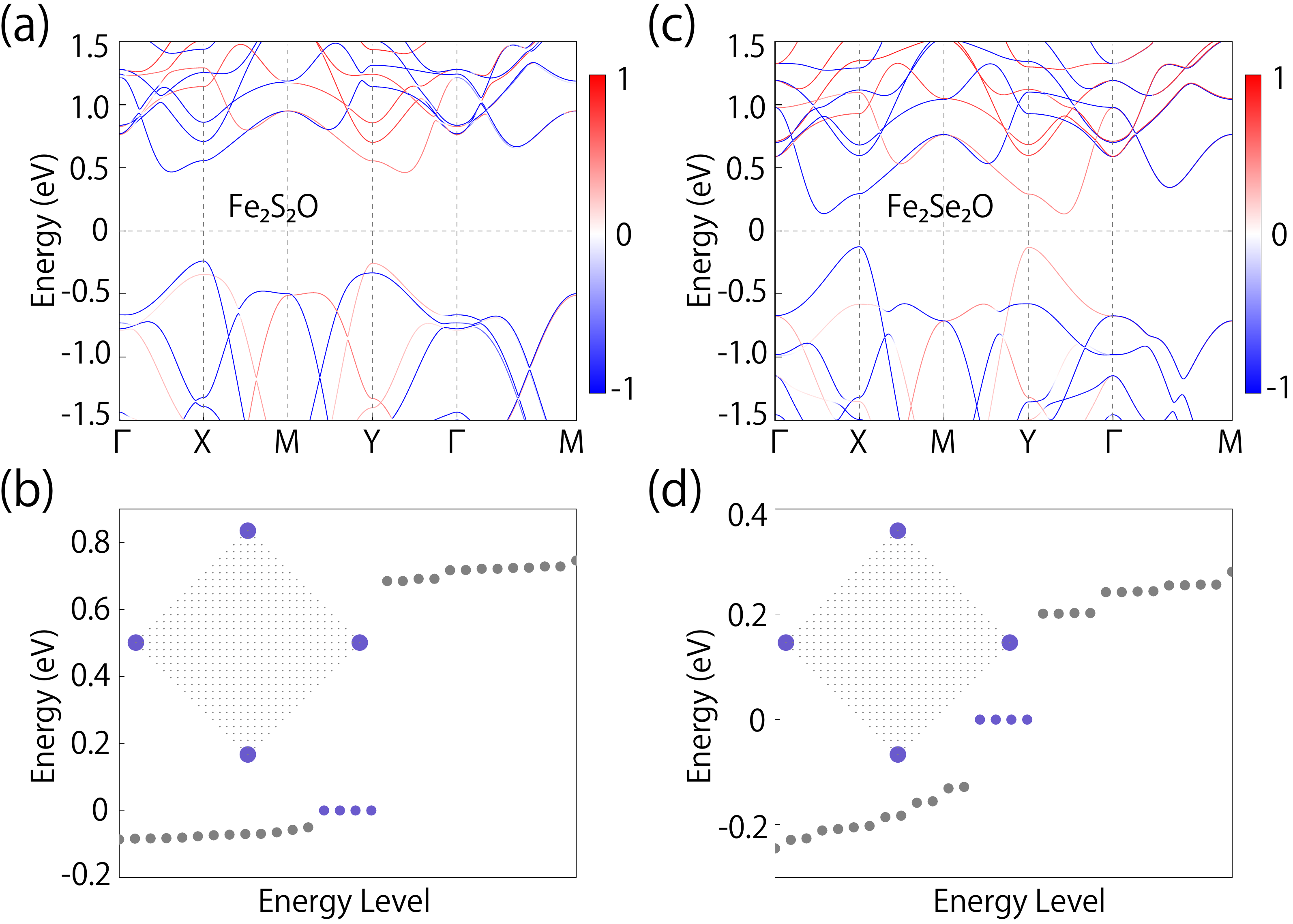}
	\caption{(a) Band structure of monolayer Fe$_2$S$_2$O including SOC, with magnetization along the $y$-axis. (c) Band structure of monolayer Fe$_2$Se$_2$O including SOC, with magnetization along the $z$-axis. (b) and (d) Energy spectra of square-shaped nanodisks of monolayer Fe$_2$S$_2$O and Fe$_2$Se$_2$O, respectively, with SOC included. Insets show the charge distributions of the zero-energy states, clearly demonstrating strong corner localization.
		\label{fig4}}
\end{figure}

As mentioned earlier, the relatively weak SOC was not taken into account in the preceding discussion. In the following, we investigate the effects of SOC on the electronic structure and topological properties of monolayer Fe\(_2\)S\(_2\)O and Fe\(_2\)Se\(_2\)O. We calculate the magnetocrystalline anisotropy energy (MAE) for the AM ground state of these materials by comparing the energies with magnetization vectors oriented along three high-symmetry directions: [010], [110], and [001], including SOC effects. Our results reveal distinct magnetic anisotropy behaviors: in Fe\(_2\)S\(_2\)O, the ground-state magnetic moment preferentially aligns along the [010] direction, with an out-of-plane vs in-plane energy difference of 798 $\mu$eV (defined as $\Delta E_{\mathrm{001-010}}$). Conversely, Fe\(_2\)Se\(_2\)O exhibits [001] alignment as its ground state, showing an energy difference of -777 $\mu$eV between out-of-plane and in-plane orientations. The resulting band structures with SOC are presented in Figs.~\ref{fig4}(a) and~\ref{fig4}(c), which reveals that the SOC has only a weak impact on the low-energy bands. This is expected, as these states predominantly originate from the S/Se atoms, where SOC is intrinsically weak. In contrast, the stronger atomic SOC arises from the Fe $3d$ orbitals, which are located far below the Fermi level and thus have limited influence on the low-energy physics. Upon introducing SOC, the two spin channels become coupled. Nevertheless, the real Chern number $\nu_R$ can still be determined through an adiabatic deformation approach. Starting from the SOC-free insulating band structure shown in Fig.~\ref{fig2}, which has $\nu_R^\uparrow = \nu_R^\downarrow = 1$, we gradually increase the SOC strength $\lambda$ from 0 to 1, with $\lambda = 1$ representing the fully interacting case. This procedure can be readily implemented in first-principles calculations. Throughout this process, the band gap remains open, indicating that the SOC-inclusive band structure is adiabatically connected to the SOC-free one. Therefore, both must possess the same topological invariant, $\nu_R^\uparrow = \nu_R^\downarrow = 1$, confirming that monolayer Fe$_2$S$_2$O and Fe$_2$Se$_2$O remain real Chern insulators in the presence of SOC. To further support this, we calculated the energy levels and corner states of square-shaped nanodisks of monolayer Fe$_2$S$_2$O and Fe$_2$Se$_2$O with SOC included, as shown in Figs.~\ref{fig4}(b) and~\ref{fig4}(d). These figures clearly demonstrate the persistence of zero-energy states localized at the corners of the nanodisks, providing further evidence for the robustness of the real Chern topological phase in the presence of SOC.

\subsection{Strain effect}
\begin{figure*}[htb]
	\centering
	\includegraphics[width=15cm]{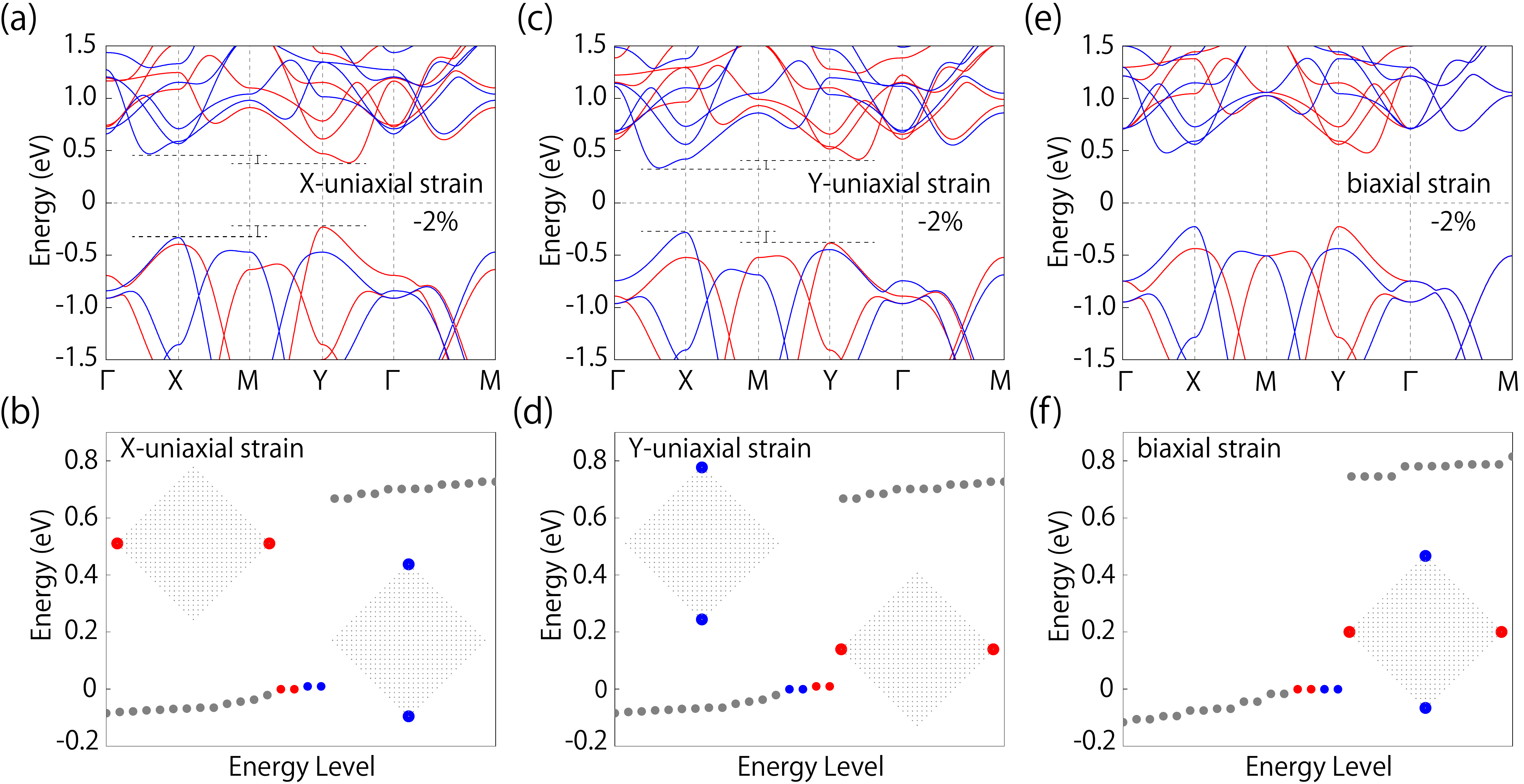}
	\caption{Band structures of monolayer Fe$_2$S$_2$O under $-2\%$ uniaxial strain along the (a) $x$- and (c) $y$-directions without SOC. (b) and (d) show the corresponding energy spectra of Fe$_2$S$_2$O nanodisks, with insets depicting charge density distributions of two sets of corner states. (e) Band structure under $-2\%$ biaxial strain and (f) the corresponding nanodisk energy spectrum. Red (blue) dots denote spin-up (spin-down) components.
		\label{fig5}}
\end{figure*}

\begin{figure*}[htb]
	\centering
	\includegraphics[width=15cm]{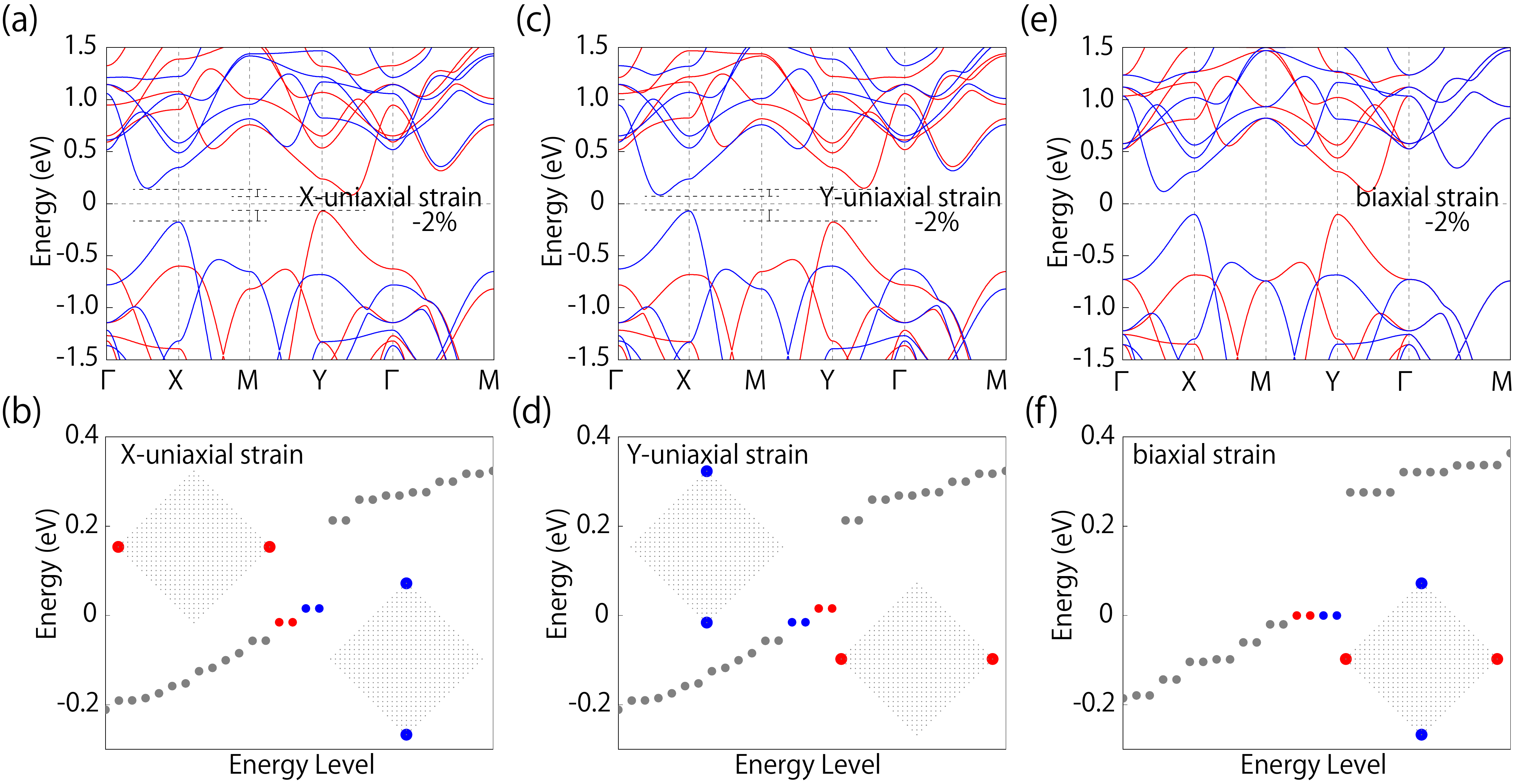}
	\caption{Band structures of monolayer Fe$_2$Se$_2$O under $-2\%$ uniaxial strain along the (a) $x$- and (c) $y$-directions without SOC. (b) and (d) show the corresponding energy spectra of Fe$_2$Se$_2$O nanodisks, with insets depicting charge density distributions of two sets of corner states. (e) Band structure under $-2\%$ biaxial strain and (f) the corresponding nanodisk energy spectrum. Red (blue) dots denote spin-up (spin-down) components.
		\label{fig6}}
\end{figure*}

In this section, we investigate the effect of strain on the electronic structures and topological properties of monolayer Fe$_2$S$_2$O and Fe$_2$Se$_2$O. We consider 2\% uniaxial and biaxial compressive strains, as shown in Fig.~\ref{fig5} and Fig.~\ref{fig6}. From Fig.~\ref{fig5} and Fig.~\ref{fig6}, one observes that under 2\% uniaxial and biaxial compressive strains, the corner states of the systems remain well-preserved and exhibit excellent robustness against mechanical deformations. This is because neither uniaxial nor biaxial strain breaks the mirror symmetry $\mathcal{M}_z$ or the inversion symmetry $\mathcal{P}$ of the system. 
It is particularly noteworthy that uniaxial strain applied along different directions (i.e., the $x$- and $y$-directions) induces distinct effects on the band structure and corner states of the system. When uniaxial strain is applied along the $x$-direction, the valley at the X point shifts downward in energy, while the valley at the Y point shifts upward, resulting in valley polarization, as shown in Figs.~\ref{fig5}(a) and~\ref{fig6}(a). This arises from the breaking of the $\mathcal{C}_{4z}\mathcal{T}$ symmetry by the uniaxial strain. Although the system remains a real Chern insulator, the topological corner states become split due to the strain: the spin-up corner states shift to lower energy, while the spin-down corner states shift to higher energy. Nevertheless, they remain well localized at the corners, as shown in the insets of Figs.~\ref{fig5}(b) and~\ref{fig6}(b). Interestingly, when uniaxial strain is applied along the $y$-direction, the valley at X shifts upward while that at Y shifts downward, reversing the valley polarization, as shown in Figs.~\ref{fig5}(c) and~\ref{fig6}(c). Correspondingly, the energy splitting of the corner states also reverses: the spin-down corner states move to lower energy, while the spin-up ones shift to higher energy (see Figs.~\ref{fig5}(d) and~\ref{fig6}(d)). Therefore, the corner states can be effectively tuned by applying uniaxial strain along different directions.

\subsection{Optical properties}
\begin{figure}[htb]
	\centering
	\includegraphics[width=8.6cm]{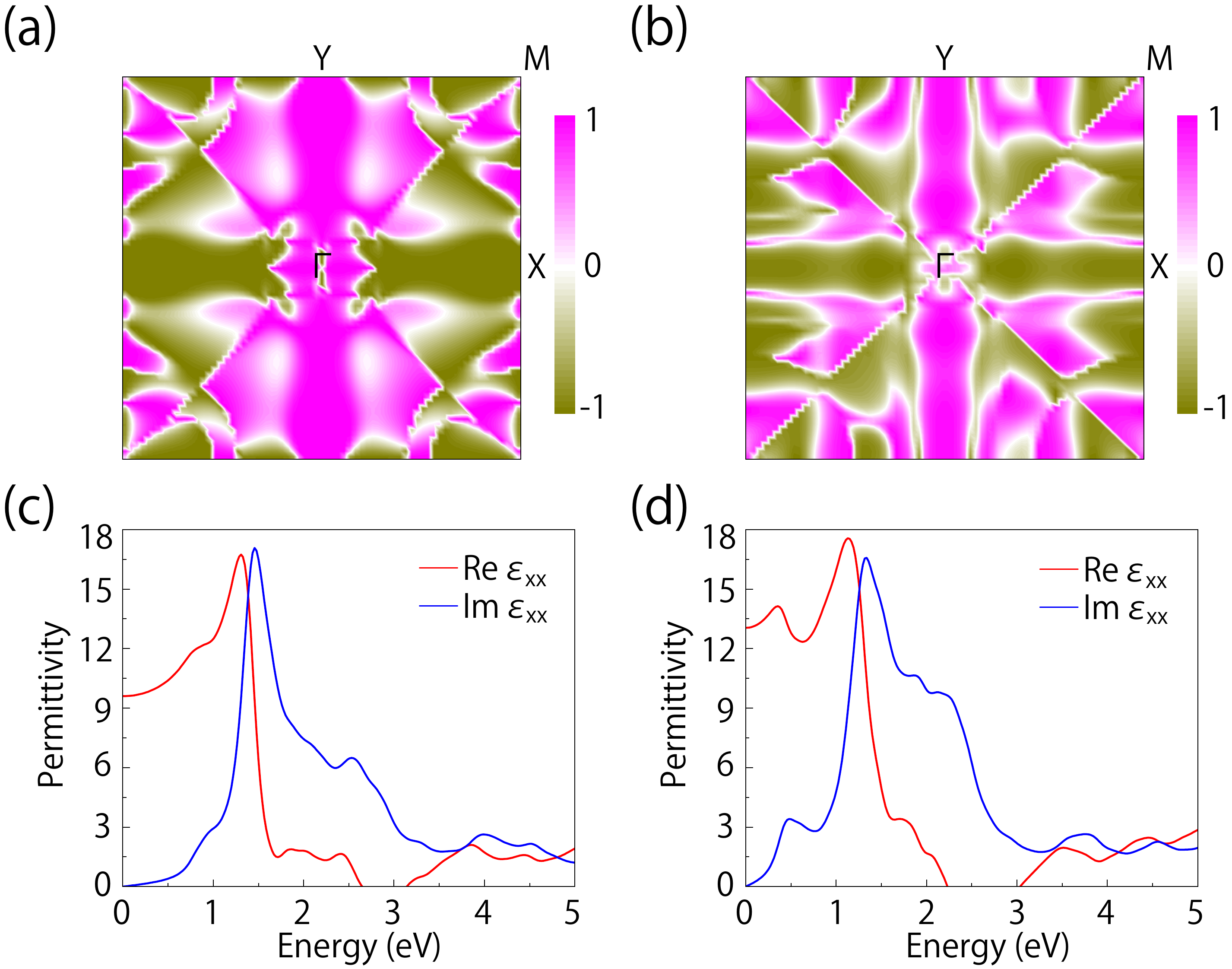}
	\caption{(a) and (b) Linear dichroism $\xi(\boldsymbol{k})$ for transitions between the highest valence band and the lowest conduction band in monolayer Fe$_2$S$_2$O and Fe$_2$Se$_2$O, respectively. (c) and (d) Real and imaginary parts of the permittivity for Fe$_2$S$_2$O and Fe$_2$Se$_2$O, respectively.		
	\label{fig7}}
\end{figure}

In this section, we investigate the optical properties of the monolayer Fe$_2$S$_2$O and Fe$_2$Se$_2$O, including linear dichroism and permittivity. Due to the presence of $\mathcal{C}_{4z}\mathcal{T}$ symmetry in the system, linear dichroism is expected to appear in the optical interband absorption. The linear dichroism $\xi^{nm}(\boldsymbol{k})$ is defined as
\begin{align}
	\begin{split}
		\xi^{nm}(\boldsymbol{k}) = \frac{|\mathcal{M}_x^{nm}(\boldsymbol{k})|^2-|\mathcal{M}_y^{nm}(\boldsymbol{k})|^2}{|\mathcal{M}_x^{nm}(\boldsymbol{k})|^2+|\mathcal{M}_y^{nm}(\boldsymbol{k})|^2}.
	\end{split}
\end{align}
where this $\boldsymbol{k}$-resolved quantity characterizes the anisotropy in the interband absorption of linearly polarized light along the $x$ and $y$ directions for interband transitions from the valence band ($m$) to the conduction band ($n$).
Here, the matrix elements given by $\mathcal{M}^{nm}_{x,y}(\boldsymbol{k})=\left\langle \psi_{\boldsymbol{k}}^n\left|\hat p_{x,y}\right|\psi_{\boldsymbol{k}}^m\right\rangle$. 
The calculated linear dichroism $\xi^{nm}(\boldsymbol{k})$ for monolayer Fe$_2$S$_2$O and Fe$_2$Se$_2$O are presented in Figs.~\ref{fig7}(a) and~\ref{fig7}(b). As shown in the Figs.~\ref{fig7}(a) and~\ref{fig7}(b), near the valley centers, the interband transitions are exclusively coupled to $x$-linearly ($y$-linearly) polarized light at the Y (X) valley. This behavior indicates that valley-selective carrier excitation can be achieved by tuning the polarization direction of the incident light.
We also computed the real and imaginary components of the permittivity for these monolayer materials. Figures~\ref{fig7}(c) and~\ref{fig7}(d) present the DFT-calculated real and imaginary parts of $\varepsilon_{xx}$, where the imaginary component, $\text{Im}(\varepsilon_{xx})$, is directly related to optical absorption. The results indicate strong optical absorption in Fe$_2$S$_2$O and Fe$_2$Se$_2$O, with intensities comparable to those of monolayer MoS$_2$ and the MoSi$_2$N$_4$ family~\cite{ahmed2015bonding,lahourpour2019structural,ermolaev2020broadband,li2020valley}.

\section{Conclusion}
In conclusion, we have identified monolayer Fe$_2$S$_2$O and Fe$_2$Se$_2$O as a new class of 2D altermagnetic real Chern insulators through first-principles calculations and theoretical analysis. These materials host altermagnetic ground states with spin-split band structures in the absence of spin–orbit coupling, and each spin channel is characterized by a nontrivial mirror real Chern number. This gives rise to symmetry-protected, spin-polarized corner states and a distinctive spin–corner coupling effect. We further demonstrate that the topological phases and their corner modes remain robust against SOC and under both uniaxial and biaxial strain. Additionally, these materials exhibit pronounced linear dichroism and strong optical absorption. Our results establish monolayer Fe$_2$S$_2$O and Fe$_2$Se$_2$O as promising platforms for exploring real Chern topology and its intriguing physical phenomena in 2D altermagnetic materials.

\bigskip
\begin{acknowledgements}
The authors thank S. A. Yang for helpful discussions. This work is supported by the National Natural Science Foundation of China (Grant No. 12204378) and the Key Program of the Natural Science Basic Research Plan of Shaanxi Province (Grant No. 2025JC-QYCX-007).
\end{acknowledgements}



\begin{thebibliography}{88}%
	\makeatletter
	\providecommand \@ifxundefined [1]{%
		\@ifx{#1\undefined}
	}%
	\providecommand \@ifnum [1]{%
		\ifnum #1\expandafter \@firstoftwo
		\else \expandafter \@secondoftwo
		\fi
	}%
	\providecommand \@ifx [1]{%
		\ifx #1\expandafter \@firstoftwo
		\else \expandafter \@secondoftwo
		\fi
	}%
	\providecommand \natexlab [1]{#1}%
	\providecommand \enquote  [1]{``#1''}%
	\providecommand \bibnamefont  [1]{#1}%
	\providecommand \bibfnamefont [1]{#1}%
	\providecommand \citenamefont [1]{#1}%
	\providecommand \href@noop [0]{\@secondoftwo}%
	\providecommand \href [0]{\begingroup \@sanitize@url \@href}%
	\providecommand \@href[1]{\@@startlink{#1}\@@href}%
	\providecommand \@@href[1]{\endgroup#1\@@endlink}%
	\providecommand \@sanitize@url [0]{\catcode `\\12\catcode `\$12\catcode
		`\&12\catcode `\#12\catcode `\^12\catcode `\_12\catcode `\%12\relax}%
	\providecommand \@@startlink[1]{}%
	\providecommand \@@endlink[0]{}%
	\providecommand \url  [0]{\begingroup\@sanitize@url \@url }%
	\providecommand \@url [1]{\endgroup\@href {#1}{\urlprefix }}%
	\providecommand \urlprefix  [0]{URL }%
	\providecommand \Eprint [0]{\href }%
	\providecommand \doibase [0]{https://doi.org/}%
	\providecommand \selectlanguage [0]{\@gobble}%
	\providecommand \bibinfo  [0]{\@secondoftwo}%
	\providecommand \bibfield  [0]{\@secondoftwo}%
	\providecommand \translation [1]{[#1]}%
	\providecommand \BibitemOpen [0]{}%
	\providecommand \bibitemStop [0]{}%
	\providecommand \bibitemNoStop [0]{.\EOS\space}%
	\providecommand \EOS [0]{\spacefactor3000\relax}%
	\providecommand \BibitemShut  [1]{\csname bibitem#1\endcsname}%
	\let\auto@bib@innerbib\@empty
	\bibitem [{\citenamefont {Hasan}\ and\ \citenamefont
		{Kane}(2010)}]{hasan2010colloquium}%
	\BibitemOpen
	\bibfield  {author} {\bibinfo {author} {\bibfnamefont {M.~Z.}\ \bibnamefont
			{Hasan}}\ and\ \bibinfo {author} {\bibfnamefont {C.~L.}\ \bibnamefont
			{Kane}},\ }\bibfield  {title} {\bibinfo {title} {Colloquium: topological
			insulators},\ }\href@noop {} {\bibfield  {journal} {\bibinfo  {journal} {Rev.
				Mod. Phys.}\ }\textbf {\bibinfo {volume} {82}},\ \bibinfo {pages} {3045}
		(\bibinfo {year} {2010})}\BibitemShut {NoStop}%
	\bibitem [{\citenamefont {Qi}\ and\ \citenamefont
		{Zhang}(2011)}]{qi2011topological}%
	\BibitemOpen
	\bibfield  {author} {\bibinfo {author} {\bibfnamefont {X.-L.}\ \bibnamefont
			{Qi}}\ and\ \bibinfo {author} {\bibfnamefont {S.-C.}\ \bibnamefont {Zhang}},\
	}\bibfield  {title} {\bibinfo {title} {Topological insulators and
			superconductors},\ }\href@noop {} {\bibfield  {journal} {\bibinfo  {journal}
			{Rev. Mod. Phys.}\ }\textbf {\bibinfo {volume} {83}},\ \bibinfo {pages}
		{1057} (\bibinfo {year} {2011})}\BibitemShut {NoStop}%
	\bibitem [{\citenamefont {Bansil}\ \emph {et~al.}(2016)\citenamefont {Bansil},
		\citenamefont {Lin},\ and\ \citenamefont {Das}}]{bansil2016colloquium}%
	\BibitemOpen
	\bibfield  {author} {\bibinfo {author} {\bibfnamefont {A.}~\bibnamefont
			{Bansil}}, \bibinfo {author} {\bibfnamefont {H.}~\bibnamefont {Lin}},\ and\
		\bibinfo {author} {\bibfnamefont {T.}~\bibnamefont {Das}},\ }\bibfield
	{title} {\bibinfo {title} {Colloquium: Topological band theory},\ }\href@noop
	{} {\bibfield  {journal} {\bibinfo  {journal} {Rev. Mod. Phys.}\ }\textbf
		{\bibinfo {volume} {88}},\ \bibinfo {pages} {021004} (\bibinfo {year}
		{2016})}\BibitemShut {NoStop}%
	\bibitem [{\citenamefont {Armitage}\ \emph {et~al.}(2018)\citenamefont
		{Armitage}, \citenamefont {Mele},\ and\ \citenamefont
		{Vishwanath}}]{armitage2018weyl}%
	\BibitemOpen
	\bibfield  {author} {\bibinfo {author} {\bibfnamefont {N.}~\bibnamefont
			{Armitage}}, \bibinfo {author} {\bibfnamefont {E.}~\bibnamefont {Mele}},\
		and\ \bibinfo {author} {\bibfnamefont {A.}~\bibnamefont {Vishwanath}},\
	}\bibfield  {title} {\bibinfo {title} {Weyl and $\mathrm{Dirac}$ semimetals
			in three-dimensional solids},\ }\href@noop {} {\bibfield  {journal} {\bibinfo
			{journal} {Rev. Mod. Phys.}\ }\textbf {\bibinfo {volume} {90}},\ \bibinfo
		{pages} {015001} (\bibinfo {year} {2018})}\BibitemShut {NoStop}%
	\bibitem [{\citenamefont {Zhao}\ \emph {et~al.}(2016)\citenamefont {Zhao},
		\citenamefont {Schnyder},\ and\ \citenamefont {Wang}}]{zhao2016unified}%
	\BibitemOpen
	\bibfield  {author} {\bibinfo {author} {\bibfnamefont {Y.}~\bibnamefont
			{Zhao}}, \bibinfo {author} {\bibfnamefont {A.~P.}\ \bibnamefont {Schnyder}},\
		and\ \bibinfo {author} {\bibfnamefont {Z.}~\bibnamefont {Wang}},\ }\bibfield
	{title} {\bibinfo {title} {Unified theory of $\mathrm{PT}$ and $\mathrm{CP}$
			invariant topological metals and nodal superconductors},\ }\href@noop {}
	{\bibfield  {journal} {\bibinfo  {journal} {Phys. Rev. Lett.}\ }\textbf
		{\bibinfo {volume} {116}},\ \bibinfo {pages} {156402} (\bibinfo {year}
		{2016})}\BibitemShut {NoStop}%
	\bibitem [{\citenamefont {Zhao}\ and\ \citenamefont {Lu}(2017)}]{zhao2017pt}%
	\BibitemOpen
	\bibfield  {author} {\bibinfo {author} {\bibfnamefont {Y.}~\bibnamefont
			{Zhao}}\ and\ \bibinfo {author} {\bibfnamefont {Y.}~\bibnamefont {Lu}},\
	}\bibfield  {title} {\bibinfo {title} {$\mathrm{PT}$-symmetric real
			$\mathrm{Dirac}$ fermions and semimetals},\ }\href@noop {} {\bibfield
		{journal} {\bibinfo  {journal} {Phys. Rev. Lett.}\ }\textbf {\bibinfo
			{volume} {118}},\ \bibinfo {pages} {056401} (\bibinfo {year}
		{2017})}\BibitemShut {NoStop}%
	\bibitem [{\citenamefont {Ahn}\ \emph {et~al.}(2018)\citenamefont {Ahn},
		\citenamefont {Kim}, \citenamefont {Kim},\ and\ \citenamefont
		{Yang}}]{ahn2018band}%
	\BibitemOpen
	\bibfield  {author} {\bibinfo {author} {\bibfnamefont {J.}~\bibnamefont
			{Ahn}}, \bibinfo {author} {\bibfnamefont {D.}~\bibnamefont {Kim}}, \bibinfo
		{author} {\bibfnamefont {Y.}~\bibnamefont {Kim}},\ and\ \bibinfo {author}
		{\bibfnamefont {B.-J.}\ \bibnamefont {Yang}},\ }\bibfield  {title} {\bibinfo
		{title} {Band topology and linking structure of nodal line semimetals with
			$\mathrm{Z}_{2}$ monopole charges},\ }\href@noop {} {\bibfield  {journal}
		{\bibinfo  {journal} {Phys. Rev. Lett.}\ }\textbf {\bibinfo {volume} {121}},\
		\bibinfo {pages} {106403} (\bibinfo {year} {2018})}\BibitemShut {NoStop}%
	\bibitem [{\citenamefont {Ahn}\ \emph {et~al.}(2019{\natexlab{a}})\citenamefont
		{Ahn}, \citenamefont {Park}, \citenamefont {Kim}, \citenamefont {Kim},\ and\
		\citenamefont {Yang}}]{ahn2019stiefel}%
	\BibitemOpen
	\bibfield  {author} {\bibinfo {author} {\bibfnamefont {J.}~\bibnamefont
			{Ahn}}, \bibinfo {author} {\bibfnamefont {S.}~\bibnamefont {Park}}, \bibinfo
		{author} {\bibfnamefont {D.}~\bibnamefont {Kim}}, \bibinfo {author}
		{\bibfnamefont {Y.}~\bibnamefont {Kim}},\ and\ \bibinfo {author}
		{\bibfnamefont {B.-J.}\ \bibnamefont {Yang}},\ }\bibfield  {title} {\bibinfo
		{title} {Stiefel--whitney classes and topological phases in band theory},\
	}\href@noop {} {\bibfield  {journal} {\bibinfo  {journal} {Chin. Phys. B}\
		}\textbf {\bibinfo {volume} {28}},\ \bibinfo {pages} {117101} (\bibinfo
		{year} {2019}{\natexlab{a}})}\BibitemShut {NoStop}%
	\bibitem [{\citenamefont {Sheng}\ \emph {et~al.}(2019)\citenamefont {Sheng},
		\citenamefont {Chen}, \citenamefont {Liu}, \citenamefont {Chen},
		\citenamefont {Yu}, \citenamefont {Zhao},\ and\ \citenamefont
		{Yang}}]{sheng2019two}%
	\BibitemOpen
	\bibfield  {author} {\bibinfo {author} {\bibfnamefont {X.-L.}\ \bibnamefont
			{Sheng}}, \bibinfo {author} {\bibfnamefont {C.}~\bibnamefont {Chen}},
		\bibinfo {author} {\bibfnamefont {H.}~\bibnamefont {Liu}}, \bibinfo {author}
		{\bibfnamefont {Z.}~\bibnamefont {Chen}}, \bibinfo {author} {\bibfnamefont
			{Z.-M.}\ \bibnamefont {Yu}}, \bibinfo {author} {\bibfnamefont
			{Y.}~\bibnamefont {Zhao}},\ and\ \bibinfo {author} {\bibfnamefont {S.~A.}\
			\bibnamefont {Yang}},\ }\bibfield  {title} {\bibinfo {title} {Two-dimensional
			second-order topological insulator in graphdiyne},\ }\href@noop {} {\bibfield
		{journal} {\bibinfo  {journal} {Phys. Rev. Lett.}\ }\textbf {\bibinfo
			{volume} {123}},\ \bibinfo {pages} {256402} (\bibinfo {year}
		{2019})}\BibitemShut {NoStop}%
	\bibitem [{\citenamefont {Lee}\ \emph {et~al.}(2020)\citenamefont {Lee},
		\citenamefont {Kim}, \citenamefont {Ahn},\ and\ \citenamefont
		{Yang}}]{lee2020two}%
	\BibitemOpen
	\bibfield  {author} {\bibinfo {author} {\bibfnamefont {E.}~\bibnamefont
			{Lee}}, \bibinfo {author} {\bibfnamefont {R.}~\bibnamefont {Kim}}, \bibinfo
		{author} {\bibfnamefont {J.}~\bibnamefont {Ahn}},\ and\ \bibinfo {author}
		{\bibfnamefont {B.-J.}\ \bibnamefont {Yang}},\ }\bibfield  {title} {\bibinfo
		{title} {Two-dimensional higher-order topology in monolayer graphdiyne},\
	}\href@noop {} {\bibfield  {journal} {\bibinfo  {journal} {npj Quantum
				Materials}\ }\textbf {\bibinfo {volume} {5}},\ \bibinfo {pages} {1} (\bibinfo
		{year} {2020})}\BibitemShut {NoStop}%
	\bibitem [{\citenamefont {Chen}\ \emph {et~al.}(2022)\citenamefont {Chen},
		\citenamefont {Zeng}, \citenamefont {Chen}, \citenamefont {Zhao},
		\citenamefont {Sheng},\ and\ \citenamefont {Yang}}]{chen2022second}%
	\BibitemOpen
	\bibfield  {author} {\bibinfo {author} {\bibfnamefont {C.}~\bibnamefont
			{Chen}}, \bibinfo {author} {\bibfnamefont {X.-T.}\ \bibnamefont {Zeng}},
		\bibinfo {author} {\bibfnamefont {Z.}~\bibnamefont {Chen}}, \bibinfo {author}
		{\bibfnamefont {Y.}~\bibnamefont {Zhao}}, \bibinfo {author} {\bibfnamefont
			{X.-L.}\ \bibnamefont {Sheng}},\ and\ \bibinfo {author} {\bibfnamefont
			{S.~A.}\ \bibnamefont {Yang}},\ }\bibfield  {title} {\bibinfo {title}
		{Second-order real nodal-line semimetal in three-dimensional graphdiyne},\
	}\href@noop {} {\bibfield  {journal} {\bibinfo  {journal} {Phys. Rev. Lett.}\
		}\textbf {\bibinfo {volume} {128}},\ \bibinfo {pages} {026405} (\bibinfo
		{year} {2022})}\BibitemShut {NoStop}%
	\bibitem [{\citenamefont {Bouhon}\ \emph {et~al.}(2020)\citenamefont {Bouhon},
		\citenamefont {Bzdu{\v{s}}ek},\ and\ \citenamefont
		{Slager}}]{bouhon2020geometric}%
	\BibitemOpen
	\bibfield  {author} {\bibinfo {author} {\bibfnamefont {A.}~\bibnamefont
			{Bouhon}}, \bibinfo {author} {\bibfnamefont {T.}~\bibnamefont
			{Bzdu{\v{s}}ek}},\ and\ \bibinfo {author} {\bibfnamefont {R.-J.}\
			\bibnamefont {Slager}},\ }\bibfield  {title} {\bibinfo {title} {Geometric
			approach to fragile topology beyond symmetry indicators},\ }\href@noop {}
	{\bibfield  {journal} {\bibinfo  {journal} {Phys. Rev. B}\ }\textbf {\bibinfo
			{volume} {102}},\ \bibinfo {pages} {115135} (\bibinfo {year}
		{2020})}\BibitemShut {NoStop}%
	\bibitem [{\citenamefont {Chen}\ \emph {et~al.}(2021)\citenamefont {Chen},
		\citenamefont {Wu}, \citenamefont {Yu}, \citenamefont {Chen}, \citenamefont
		{Zhao}, \citenamefont {Sheng},\ and\ \citenamefont
		{Yang}}]{chen2021graphyne}%
	\BibitemOpen
	\bibfield  {author} {\bibinfo {author} {\bibfnamefont {C.}~\bibnamefont
			{Chen}}, \bibinfo {author} {\bibfnamefont {W.}~\bibnamefont {Wu}}, \bibinfo
		{author} {\bibfnamefont {Z.-M.}\ \bibnamefont {Yu}}, \bibinfo {author}
		{\bibfnamefont {Z.}~\bibnamefont {Chen}}, \bibinfo {author} {\bibfnamefont
			{Y.}~\bibnamefont {Zhao}}, \bibinfo {author} {\bibfnamefont {X.-L.}\
			\bibnamefont {Sheng}},\ and\ \bibinfo {author} {\bibfnamefont {S.~A.}\
			\bibnamefont {Yang}},\ }\bibfield  {title} {\bibinfo {title} {Graphyne as a
			second-order and real chern topological insulator in two dimensions},\
	}\href@noop {} {\bibfield  {journal} {\bibinfo  {journal} {Phys. Rev. B}\
		}\textbf {\bibinfo {volume} {104}},\ \bibinfo {pages} {085205} (\bibinfo
		{year} {2021})}\BibitemShut {NoStop}%
	\bibitem [{\citenamefont {Qian}\ \emph {et~al.}(2021)\citenamefont {Qian},
		\citenamefont {Liu},\ and\ \citenamefont {Yao}}]{qian2021second}%
	\BibitemOpen
	\bibfield  {author} {\bibinfo {author} {\bibfnamefont {S.}~\bibnamefont
			{Qian}}, \bibinfo {author} {\bibfnamefont {C.-C.}\ \bibnamefont {Liu}},\ and\
		\bibinfo {author} {\bibfnamefont {Y.}~\bibnamefont {Yao}},\ }\bibfield
	{title} {\bibinfo {title} {Second-order topological insulator state in
			hexagonal lattices and its abundant material candidates},\ }\href@noop {}
	{\bibfield  {journal} {\bibinfo  {journal} {Phys. Rev. B}\ }\textbf {\bibinfo
			{volume} {104}},\ \bibinfo {pages} {245427} (\bibinfo {year}
		{2021})}\BibitemShut {NoStop}%
	\bibitem [{\citenamefont {Pan}\ \emph {et~al.}(2022)\citenamefont {Pan},
		\citenamefont {Li}, \citenamefont {Fan},\ and\ \citenamefont
		{Huang}}]{pan2022two}%
	\BibitemOpen
	\bibfield  {author} {\bibinfo {author} {\bibfnamefont {M.}~\bibnamefont
			{Pan}}, \bibinfo {author} {\bibfnamefont {D.}~\bibnamefont {Li}}, \bibinfo
		{author} {\bibfnamefont {J.}~\bibnamefont {Fan}},\ and\ \bibinfo {author}
		{\bibfnamefont {H.}~\bibnamefont {Huang}},\ }\bibfield  {title} {\bibinfo
		{title} {Two-dimensional stiefel-whitney insulators in liganded xenes},\
	}\href@noop {} {\bibfield  {journal} {\bibinfo  {journal} {npj Comput.
				Mater.}\ }\textbf {\bibinfo {volume} {8}},\ \bibinfo {pages} {1} (\bibinfo
		{year} {2022})}\BibitemShut {NoStop}%
	\bibitem [{\citenamefont {Guo}\ \emph {et~al.}(2022)\citenamefont {Guo},
		\citenamefont {Deng}, \citenamefont {Xie},\ and\ \citenamefont
		{Wang}}]{guo2022quadrupole}%
	\BibitemOpen
	\bibfield  {author} {\bibinfo {author} {\bibfnamefont {Z.}~\bibnamefont
			{Guo}}, \bibinfo {author} {\bibfnamefont {J.}~\bibnamefont {Deng}}, \bibinfo
		{author} {\bibfnamefont {Y.}~\bibnamefont {Xie}},\ and\ \bibinfo {author}
		{\bibfnamefont {Z.}~\bibnamefont {Wang}},\ }\bibfield  {title} {\bibinfo
		{title} {Quadrupole topological insulators in $\mathrm{Ta_{2}M_{3}Te_{5}}
			(\mathrm{M} = \mathrm{Ni, Pd})$ monolayers},\ }\href@noop {} {\bibfield
		{journal} {\bibinfo  {journal} {npj Quantum Materials}\ }\textbf {\bibinfo
			{volume} {7}},\ \bibinfo {pages} {87} (\bibinfo {year} {2022})}\BibitemShut
	{NoStop}%
	\bibitem [{\citenamefont {Park}\ \emph {et~al.}(2019)\citenamefont {Park},
		\citenamefont {Kim}, \citenamefont {Cho},\ and\ \citenamefont
		{Lee}}]{park2019higher}%
	\BibitemOpen
	\bibfield  {author} {\bibinfo {author} {\bibfnamefont {M.~J.}\ \bibnamefont
			{Park}}, \bibinfo {author} {\bibfnamefont {Y.}~\bibnamefont {Kim}}, \bibinfo
		{author} {\bibfnamefont {G.~Y.}\ \bibnamefont {Cho}},\ and\ \bibinfo {author}
		{\bibfnamefont {S.}~\bibnamefont {Lee}},\ }\bibfield  {title} {\bibinfo
		{title} {Higher-order topological insulator in twisted bilayer graphene},\
	}\href@noop {} {\bibfield  {journal} {\bibinfo  {journal} {Phys. Rev. Lett.}\
		}\textbf {\bibinfo {volume} {123}},\ \bibinfo {pages} {216803} (\bibinfo
		{year} {2019})}\BibitemShut {NoStop}%
	\bibitem [{\citenamefont {Ahn}\ \emph {et~al.}(2019{\natexlab{b}})\citenamefont
		{Ahn}, \citenamefont {Park},\ and\ \citenamefont {Yang}}]{ahn2019failure}%
	\BibitemOpen
	\bibfield  {author} {\bibinfo {author} {\bibfnamefont {J.}~\bibnamefont
			{Ahn}}, \bibinfo {author} {\bibfnamefont {S.}~\bibnamefont {Park}},\ and\
		\bibinfo {author} {\bibfnamefont {B.-J.}\ \bibnamefont {Yang}},\ }\bibfield
	{title} {\bibinfo {title} {Failure of nielsen-ninomiya theorem and fragile
			topology in two-dimensional systems with space-time inversion symmetry:
			application to twisted bilayer graphene at magic angle},\ }\href@noop {}
	{\bibfield  {journal} {\bibinfo  {journal} {Phys. Rev. X}\ }\textbf {\bibinfo
			{volume} {9}},\ \bibinfo {pages} {021013} (\bibinfo {year}
		{2019}{\natexlab{b}})}\BibitemShut {NoStop}%
	\bibitem [{\citenamefont {Han}\ \emph {et~al.}(2024)\citenamefont {Han},
		\citenamefont {Cui}, \citenamefont {Li}, \citenamefont {Zhang}, \citenamefont
		{Zhang}, \citenamefont {Yu},\ and\ \citenamefont
		{Yao}}]{han2024cornertronics}%
	\BibitemOpen
	\bibfield  {author} {\bibinfo {author} {\bibfnamefont {Y.}~\bibnamefont
			{Han}}, \bibinfo {author} {\bibfnamefont {C.}~\bibnamefont {Cui}}, \bibinfo
		{author} {\bibfnamefont {X.-P.}\ \bibnamefont {Li}}, \bibinfo {author}
		{\bibfnamefont {T.-T.}\ \bibnamefont {Zhang}}, \bibinfo {author}
		{\bibfnamefont {Z.}~\bibnamefont {Zhang}}, \bibinfo {author} {\bibfnamefont
			{Z.-M.}\ \bibnamefont {Yu}},\ and\ \bibinfo {author} {\bibfnamefont
			{Y.}~\bibnamefont {Yao}},\ }\bibfield  {title} {\bibinfo {title}
		{Cornertronics in two-dimensional second-order topological insulators},\
	}\href@noop {} {\bibfield  {journal} {\bibinfo  {journal} {Phys. Rev. Lett.}\
		}\textbf {\bibinfo {volume} {133}},\ \bibinfo {pages} {176602} (\bibinfo
		{year} {2024})}\BibitemShut {NoStop}%
	\bibitem [{\citenamefont {Zhang}\ \emph {et~al.}(2023)\citenamefont {Zhang},
		\citenamefont {He}, \citenamefont {Liu}, \citenamefont {Dai}, \citenamefont
		{Liu}, \citenamefont {Chen}, \citenamefont {Wu}, \citenamefont {Zhu},\ and\
		\citenamefont {Yang}}]{zhang2023magnetic}%
	\BibitemOpen
	\bibfield  {author} {\bibinfo {author} {\bibfnamefont {X.}~\bibnamefont
			{Zhang}}, \bibinfo {author} {\bibfnamefont {T.}~\bibnamefont {He}}, \bibinfo
		{author} {\bibfnamefont {Y.}~\bibnamefont {Liu}}, \bibinfo {author}
		{\bibfnamefont {X.}~\bibnamefont {Dai}}, \bibinfo {author} {\bibfnamefont
			{G.}~\bibnamefont {Liu}}, \bibinfo {author} {\bibfnamefont {C.}~\bibnamefont
			{Chen}}, \bibinfo {author} {\bibfnamefont {W.}~\bibnamefont {Wu}}, \bibinfo
		{author} {\bibfnamefont {J.}~\bibnamefont {Zhu}},\ and\ \bibinfo {author}
		{\bibfnamefont {S.~A.}\ \bibnamefont {Yang}},\ }\bibfield  {title} {\bibinfo
		{title} {Magnetic real chern insulator in 2d metal--organic frameworks},\
	}\href@noop {} {\bibfield  {journal} {\bibinfo  {journal} {Nano Lett.}\
		}\textbf {\bibinfo {volume} {23}},\ \bibinfo {pages} {7358} (\bibinfo {year}
		{2023})}\BibitemShut {NoStop}%
	\bibitem [{\citenamefont {Naka}\ \emph {et~al.}(2019)\citenamefont {Naka},
		\citenamefont {Hayami}, \citenamefont {Kusunose}, \citenamefont {Yanagi},
		\citenamefont {Motome},\ and\ \citenamefont {Seo}}]{Naka19NC}%
	\BibitemOpen
	\bibfield  {author} {\bibinfo {author} {\bibfnamefont {M.}~\bibnamefont
			{Naka}}, \bibinfo {author} {\bibfnamefont {S.}~\bibnamefont {Hayami}},
		\bibinfo {author} {\bibfnamefont {H.}~\bibnamefont {Kusunose}}, \bibinfo
		{author} {\bibfnamefont {Y.}~\bibnamefont {Yanagi}}, \bibinfo {author}
		{\bibfnamefont {Y.}~\bibnamefont {Motome}},\ and\ \bibinfo {author}
		{\bibfnamefont {H.}~\bibnamefont {Seo}},\ }\bibfield  {title} {\bibinfo
		{title} {{Spin current generation in organic antiferromagnets}},\ }\href@noop
	{} {\bibfield  {journal} {\bibinfo  {journal} {Nat. Commun.}\ }\textbf
		{\bibinfo {volume} {10}},\ \bibinfo {pages} {4305} (\bibinfo {year}
		{2019})}\BibitemShut {NoStop}%
	\bibitem [{\citenamefont {Ahn}\ \emph {et~al.}(2019{\natexlab{c}})\citenamefont
		{Ahn}, \citenamefont {Hariki}, \citenamefont {Lee},\ and\ \citenamefont
		{Kune\ifmmode~\check{s}\else \v{s}\fi{}}}]{Ahn19PRB}%
	\BibitemOpen
	\bibfield  {author} {\bibinfo {author} {\bibfnamefont {K.-H.}\ \bibnamefont
			{Ahn}}, \bibinfo {author} {\bibfnamefont {A.}~\bibnamefont {Hariki}},
		\bibinfo {author} {\bibfnamefont {K.-W.}\ \bibnamefont {Lee}},\ and\ \bibinfo
		{author} {\bibfnamefont {J.}~\bibnamefont {Kune\ifmmode~\check{s}\else
				\v{s}\fi{}}},\ }\bibfield  {title} {\bibinfo {title} {{Antiferromagnetism in
				${\text{RuO}}_{2}$ as $d$-wave Pomeranchuk instability}},\ }\href@noop {}
	{\bibfield  {journal} {\bibinfo  {journal} {Phys. Rev. B}\ }\textbf {\bibinfo
			{volume} {99}},\ \bibinfo {pages} {184432} (\bibinfo {year}
		{2019}{\natexlab{c}})}\BibitemShut {NoStop}%
	\bibitem [{\citenamefont {Hayami}\ \emph {et~al.}(2020)\citenamefont {Hayami},
		\citenamefont {Yanagi},\ and\ \citenamefont {Kusunose}}]{Hayami20PRB}%
	\BibitemOpen
	\bibfield  {author} {\bibinfo {author} {\bibfnamefont {S.}~\bibnamefont
			{Hayami}}, \bibinfo {author} {\bibfnamefont {Y.}~\bibnamefont {Yanagi}},\
		and\ \bibinfo {author} {\bibfnamefont {H.}~\bibnamefont {Kusunose}},\
	}\bibfield  {title} {\bibinfo {title} {{Bottom-up design of spin-split and
				reshaped electronic band structures in antiferromagnets without spin-orbit
				coupling: Procedure on the basis of augmented multipoles}},\ }\href@noop {}
	{\bibfield  {journal} {\bibinfo  {journal} {Phys. Rev. B}\ }\textbf {\bibinfo
			{volume} {102}},\ \bibinfo {pages} {144441} (\bibinfo {year}
		{2020})}\BibitemShut {NoStop}%
	\bibitem [{\citenamefont {Yuan}\ \emph {et~al.}(2020)\citenamefont {Yuan},
		\citenamefont {Wang}, \citenamefont {Luo}, \citenamefont {Rashba},\ and\
		\citenamefont {Zunger}}]{yuanLD20PRB}%
	\BibitemOpen
	\bibfield  {author} {\bibinfo {author} {\bibfnamefont {L.-D.}\ \bibnamefont
			{Yuan}}, \bibinfo {author} {\bibfnamefont {Z.}~\bibnamefont {Wang}}, \bibinfo
		{author} {\bibfnamefont {J.-W.}\ \bibnamefont {Luo}}, \bibinfo {author}
		{\bibfnamefont {E.~I.}\ \bibnamefont {Rashba}},\ and\ \bibinfo {author}
		{\bibfnamefont {A.}~\bibnamefont {Zunger}},\ }\bibfield  {title} {\bibinfo
		{title} {{Giant momentum-dependent spin splitting in centrosymmetric low-$Z$
				antiferromagnets}},\ }\href@noop {} {\bibfield  {journal} {\bibinfo
			{journal} {Phys. Rev. B}\ }\textbf {\bibinfo {volume} {102}},\ \bibinfo
		{pages} {014422} (\bibinfo {year} {2020})}\BibitemShut {NoStop}%
	\bibitem [{\citenamefont {{\v{S}}mejkal}\ \emph {et~al.}(2020)\citenamefont
		{{\v{S}}mejkal}, \citenamefont {Gonz{\'a}lez-Hern{\'a}ndez}, \citenamefont
		{Jungwirth},\ and\ \citenamefont {Sinova}}]{Smejkal2020}%
	\BibitemOpen
	\bibfield  {author} {\bibinfo {author} {\bibfnamefont {L.}~\bibnamefont
			{{\v{S}}mejkal}}, \bibinfo {author} {\bibfnamefont {R.}~\bibnamefont
			{Gonz{\'a}lez-Hern{\'a}ndez}}, \bibinfo {author} {\bibfnamefont
			{T.}~\bibnamefont {Jungwirth}},\ and\ \bibinfo {author} {\bibfnamefont
			{J.}~\bibnamefont {Sinova}},\ }\bibfield  {title} {\bibinfo {title} {Crystal
			time-reversal symmetry breaking and spontaneous hall effect in collinear
			antiferromagnets},\ }\href@noop {} {\bibfield  {journal} {\bibinfo  {journal}
			{Sci. Adv.}\ }\textbf {\bibinfo {volume} {6}},\ \bibinfo {pages} {eaaz8809}
		(\bibinfo {year} {2020})}\BibitemShut {NoStop}%
	\bibitem [{\citenamefont {Mazin}\ \emph {et~al.}(2021)\citenamefont {Mazin},
		\citenamefont {Koepernik}, \citenamefont {Johannes}, \citenamefont
		{Gonz{\'a}lez-Hern{\'a}ndez},\ and\ \citenamefont
		{{\v{S}}mejkal}}]{Mazin21PNAS}%
	\BibitemOpen
	\bibfield  {author} {\bibinfo {author} {\bibfnamefont {I.~I.}\ \bibnamefont
			{Mazin}}, \bibinfo {author} {\bibfnamefont {K.}~\bibnamefont {Koepernik}},
		\bibinfo {author} {\bibfnamefont {M.~D.}\ \bibnamefont {Johannes}}, \bibinfo
		{author} {\bibfnamefont {R.}~\bibnamefont {Gonz{\'a}lez-Hern{\'a}ndez}},\
		and\ \bibinfo {author} {\bibfnamefont {L.}~\bibnamefont {{\v{S}}mejkal}},\
	}\bibfield  {title} {\bibinfo {title} {{Prediction of unconventional
				magnetism in doped $\text{FeSb}_{2}$ }},\ }\href@noop {} {\bibfield
		{journal} {\bibinfo  {journal} {P. Natl. Acad. Sci. Usa.}\ }\textbf {\bibinfo
			{volume} {118}},\ \bibinfo {pages} {e2108924118} (\bibinfo {year}
		{2021})}\BibitemShut {NoStop}%
	\bibitem [{\citenamefont {{\v{S}}mejkal}\ \emph
		{et~al.}(2022{\natexlab{a}})\citenamefont {{\v{S}}mejkal}, \citenamefont
		{Sinova},\ and\ \citenamefont {Jungwirth}}]{vsmejkal2022beyond}%
	\BibitemOpen
	\bibfield  {author} {\bibinfo {author} {\bibfnamefont {L.}~\bibnamefont
			{{\v{S}}mejkal}}, \bibinfo {author} {\bibfnamefont {J.}~\bibnamefont
			{Sinova}},\ and\ \bibinfo {author} {\bibfnamefont {T.}~\bibnamefont
			{Jungwirth}},\ }\bibfield  {title} {\bibinfo {title} {Beyond conventional
			ferromagnetism and antiferromagnetism: A phase with nonrelativistic spin and
			crystal rotation symmetry},\ }\href@noop {} {\bibfield  {journal} {\bibinfo
			{journal} {Phys. Rev. X}\ }\textbf {\bibinfo {volume} {12}},\ \bibinfo
		{pages} {031042} (\bibinfo {year} {2022}{\natexlab{a}})}\BibitemShut
	{NoStop}%
	\bibitem [{\citenamefont {{\v{S}}mejkal}\ \emph
		{et~al.}(2022{\natexlab{b}})\citenamefont {{\v{S}}mejkal}, \citenamefont
		{Sinova},\ and\ \citenamefont {Jungwirth}}]{vsmejkal2022emerging}%
	\BibitemOpen
	\bibfield  {author} {\bibinfo {author} {\bibfnamefont {L.}~\bibnamefont
			{{\v{S}}mejkal}}, \bibinfo {author} {\bibfnamefont {J.}~\bibnamefont
			{Sinova}},\ and\ \bibinfo {author} {\bibfnamefont {T.}~\bibnamefont
			{Jungwirth}},\ }\bibfield  {title} {\bibinfo {title} {Emerging research
			landscape of altermagnetism},\ }\href@noop {} {\bibfield  {journal} {\bibinfo
			{journal} {Phys. Rev. X}\ }\textbf {\bibinfo {volume} {12}},\ \bibinfo
		{pages} {040501} (\bibinfo {year} {2022}{\natexlab{b}})}\BibitemShut
	{NoStop}%
	\bibitem [{\citenamefont {Bai}\ \emph {et~al.}(2024)\citenamefont {Bai},
		\citenamefont {Feng}, \citenamefont {Liu}, \citenamefont {{\v{S}}mejkal},
		\citenamefont {Mokrousov},\ and\ \citenamefont
		{Yao}}]{bai2024altermagnetism}%
	\BibitemOpen
	\bibfield  {author} {\bibinfo {author} {\bibfnamefont {L.}~\bibnamefont
			{Bai}}, \bibinfo {author} {\bibfnamefont {W.}~\bibnamefont {Feng}}, \bibinfo
		{author} {\bibfnamefont {S.}~\bibnamefont {Liu}}, \bibinfo {author}
		{\bibfnamefont {L.}~\bibnamefont {{\v{S}}mejkal}}, \bibinfo {author}
		{\bibfnamefont {Y.}~\bibnamefont {Mokrousov}},\ and\ \bibinfo {author}
		{\bibfnamefont {Y.}~\bibnamefont {Yao}},\ }\bibfield  {title} {\bibinfo
		{title} {Altermagnetism: Exploring new frontiers in magnetism and
			spintronics},\ }\href@noop {} {\bibfield  {journal} {\bibinfo  {journal}
			{Adv. Funct. Mater.}\ }\textbf {\bibinfo {volume} {34}},\ \bibinfo {pages}
		{2409327} (\bibinfo {year} {2024})}\BibitemShut {NoStop}%
	\bibitem [{\citenamefont {Fender}\ \emph {et~al.}(2025)\citenamefont {Fender},
		\citenamefont {Gonzalez},\ and\ \citenamefont
		{Bediako}}]{fender2025altermagnetism}%
	\BibitemOpen
	\bibfield  {author} {\bibinfo {author} {\bibfnamefont {S.~S.}\ \bibnamefont
			{Fender}}, \bibinfo {author} {\bibfnamefont {O.}~\bibnamefont {Gonzalez}},\
		and\ \bibinfo {author} {\bibfnamefont {D.~K.}\ \bibnamefont {Bediako}},\
	}\bibfield  {title} {\bibinfo {title} {Altermagnetism: A chemical
			perspective},\ }\href@noop {} {\bibfield  {journal} {\bibinfo  {journal} {J.
				Am. Chem. Soc.}\ }\textbf {\bibinfo {volume} {147}},\ \bibinfo {pages} {2257}
		(\bibinfo {year} {2025})}\BibitemShut {NoStop}%
	\bibitem [{\citenamefont {Liu}\ \emph {et~al.}(2022)\citenamefont {Liu},
		\citenamefont {Li}, \citenamefont {Han}, \citenamefont {Wan},\ and\
		\citenamefont {Liu}}]{liu2022spin}%
	\BibitemOpen
	\bibfield  {author} {\bibinfo {author} {\bibfnamefont {P.}~\bibnamefont
			{Liu}}, \bibinfo {author} {\bibfnamefont {J.}~\bibnamefont {Li}}, \bibinfo
		{author} {\bibfnamefont {J.}~\bibnamefont {Han}}, \bibinfo {author}
		{\bibfnamefont {X.}~\bibnamefont {Wan}},\ and\ \bibinfo {author}
		{\bibfnamefont {Q.}~\bibnamefont {Liu}},\ }\bibfield  {title} {\bibinfo
		{title} {Spin-group symmetry in magnetic materials with negligible spin-orbit
			coupling},\ }\href@noop {} {\bibfield  {journal} {\bibinfo  {journal} {Phys.
				Rev. X}\ }\textbf {\bibinfo {volume} {12}},\ \bibinfo {pages} {021016}
		(\bibinfo {year} {2022})}\BibitemShut {NoStop}%
	\bibitem [{\citenamefont {Xiao}\ \emph {et~al.}(2024)\citenamefont {Xiao},
		\citenamefont {Zhao}, \citenamefont {Li}, \citenamefont {Shindou},\ and\
		\citenamefont {Song}}]{xiao2024spin}%
	\BibitemOpen
	\bibfield  {author} {\bibinfo {author} {\bibfnamefont {Z.}~\bibnamefont
			{Xiao}}, \bibinfo {author} {\bibfnamefont {J.}~\bibnamefont {Zhao}}, \bibinfo
		{author} {\bibfnamefont {Y.}~\bibnamefont {Li}}, \bibinfo {author}
		{\bibfnamefont {R.}~\bibnamefont {Shindou}},\ and\ \bibinfo {author}
		{\bibfnamefont {Z.-D.}\ \bibnamefont {Song}},\ }\bibfield  {title} {\bibinfo
		{title} {Spin space groups: Full classification and applications},\
	}\href@noop {} {\bibfield  {journal} {\bibinfo  {journal} {Phys. Rev. X}\
		}\textbf {\bibinfo {volume} {14}},\ \bibinfo {pages} {031037} (\bibinfo
		{year} {2024})}\BibitemShut {NoStop}%
	\bibitem [{\citenamefont {Chen}\ \emph {et~al.}(2024)\citenamefont {Chen},
		\citenamefont {Ren}, \citenamefont {Zhu}, \citenamefont {Yu}, \citenamefont
		{Zhang}, \citenamefont {Liu}, \citenamefont {Li}, \citenamefont {Liu},
		\citenamefont {Li},\ and\ \citenamefont {Liu}}]{chen2024enumeration}%
	\BibitemOpen
	\bibfield  {author} {\bibinfo {author} {\bibfnamefont {X.}~\bibnamefont
			{Chen}}, \bibinfo {author} {\bibfnamefont {J.}~\bibnamefont {Ren}}, \bibinfo
		{author} {\bibfnamefont {Y.}~\bibnamefont {Zhu}}, \bibinfo {author}
		{\bibfnamefont {Y.}~\bibnamefont {Yu}}, \bibinfo {author} {\bibfnamefont
			{A.}~\bibnamefont {Zhang}}, \bibinfo {author} {\bibfnamefont
			{P.}~\bibnamefont {Liu}}, \bibinfo {author} {\bibfnamefont {J.}~\bibnamefont
			{Li}}, \bibinfo {author} {\bibfnamefont {Y.}~\bibnamefont {Liu}}, \bibinfo
		{author} {\bibfnamefont {C.}~\bibnamefont {Li}},\ and\ \bibinfo {author}
		{\bibfnamefont {Q.}~\bibnamefont {Liu}},\ }\bibfield  {title} {\bibinfo
		{title} {Enumeration and representation theory of spin space groups},\
	}\href@noop {} {\bibfield  {journal} {\bibinfo  {journal} {Phys. Rev. X}\
		}\textbf {\bibinfo {volume} {14}},\ \bibinfo {pages} {031038} (\bibinfo
		{year} {2024})}\BibitemShut {NoStop}%
	\bibitem [{\citenamefont {Jiang}\ \emph {et~al.}(2024)\citenamefont {Jiang},
		\citenamefont {Song}, \citenamefont {Zhu}, \citenamefont {Fang},
		\citenamefont {Weng}, \citenamefont {Liu}, \citenamefont {Yang},\ and\
		\citenamefont {Fang}}]{jiang2024enumeration}%
	\BibitemOpen
	\bibfield  {author} {\bibinfo {author} {\bibfnamefont {Y.}~\bibnamefont
			{Jiang}}, \bibinfo {author} {\bibfnamefont {Z.}~\bibnamefont {Song}},
		\bibinfo {author} {\bibfnamefont {T.}~\bibnamefont {Zhu}}, \bibinfo {author}
		{\bibfnamefont {Z.}~\bibnamefont {Fang}}, \bibinfo {author} {\bibfnamefont
			{H.}~\bibnamefont {Weng}}, \bibinfo {author} {\bibfnamefont {Z.-X.}\
			\bibnamefont {Liu}}, \bibinfo {author} {\bibfnamefont {J.}~\bibnamefont
			{Yang}},\ and\ \bibinfo {author} {\bibfnamefont {C.}~\bibnamefont {Fang}},\
	}\bibfield  {title} {\bibinfo {title} {Enumeration of spin-space groups:
			Toward a complete description of symmetries of magnetic orders},\ }\href@noop
	{} {\bibfield  {journal} {\bibinfo  {journal} {Phys. Rev. X}\ }\textbf
		{\bibinfo {volume} {14}},\ \bibinfo {pages} {031039} (\bibinfo {year}
		{2024})}\BibitemShut {NoStop}%
	\bibitem [{\citenamefont {Feng}\ \emph {et~al.}(2022)\citenamefont {Feng},
		\citenamefont {Zhou}, \citenamefont {{\v{S}}mejkal}, \citenamefont {Wu},
		\citenamefont {Zhu}, \citenamefont {Guo}, \citenamefont
		{Gonz{\'a}lez-Hern{\'a}ndez}, \citenamefont {Wang}, \citenamefont {Yan},
		\citenamefont {Qin} \emph {et~al.}}]{feng2022anomalous}%
	\BibitemOpen
	\bibfield  {author} {\bibinfo {author} {\bibfnamefont {Z.}~\bibnamefont
			{Feng}}, \bibinfo {author} {\bibfnamefont {X.}~\bibnamefont {Zhou}}, \bibinfo
		{author} {\bibfnamefont {L.}~\bibnamefont {{\v{S}}mejkal}}, \bibinfo {author}
		{\bibfnamefont {L.}~\bibnamefont {Wu}}, \bibinfo {author} {\bibfnamefont
			{Z.}~\bibnamefont {Zhu}}, \bibinfo {author} {\bibfnamefont {H.}~\bibnamefont
			{Guo}}, \bibinfo {author} {\bibfnamefont {R.}~\bibnamefont
			{Gonz{\'a}lez-Hern{\'a}ndez}}, \bibinfo {author} {\bibfnamefont
			{X.}~\bibnamefont {Wang}}, \bibinfo {author} {\bibfnamefont {H.}~\bibnamefont
			{Yan}}, \bibinfo {author} {\bibfnamefont {P.}~\bibnamefont {Qin}}, \emph
		{et~al.},\ }\bibfield  {title} {\bibinfo {title} {An anomalous hall effect in
			altermagnetic ruthenium dioxide},\ }\href@noop {} {\bibfield  {journal}
		{\bibinfo  {journal} {Nat. Electron.}\ }\textbf {\bibinfo {volume} {5}},\
		\bibinfo {pages} {735} (\bibinfo {year} {2022})}\BibitemShut {NoStop}%
	\bibitem [{\citenamefont {Wu}\ \emph {et~al.}(2007)\citenamefont {Wu},
		\citenamefont {Sun}, \citenamefont {Fradkin},\ and\ \citenamefont
		{Zhang}}]{CJWu07PRB}%
	\BibitemOpen
	\bibfield  {author} {\bibinfo {author} {\bibfnamefont {C.}~\bibnamefont
			{Wu}}, \bibinfo {author} {\bibfnamefont {K.}~\bibnamefont {Sun}}, \bibinfo
		{author} {\bibfnamefont {E.}~\bibnamefont {Fradkin}},\ and\ \bibinfo {author}
		{\bibfnamefont {S.-C.}\ \bibnamefont {Zhang}},\ }\bibfield  {title} {\bibinfo
		{title} {{Fermi} liquid instabilities in the spin channel},\ }\href@noop {}
	{\bibfield  {journal} {\bibinfo  {journal} {Phys. Rev. B}\ }\textbf {\bibinfo
			{volume} {75}},\ \bibinfo {pages} {115103} (\bibinfo {year}
		{2007})}\BibitemShut {NoStop}%
	\bibitem [{\citenamefont {Ma}\ \emph {et~al.}(2021)\citenamefont {Ma},
		\citenamefont {Hu}, \citenamefont {Li}, \citenamefont {Liu}, \citenamefont
		{Yao}, \citenamefont {Jia},\ and\ \citenamefont
		{Liu}}]{ma2021multifunctional}%
	\BibitemOpen
	\bibfield  {author} {\bibinfo {author} {\bibfnamefont {H.-Y.}\ \bibnamefont
			{Ma}}, \bibinfo {author} {\bibfnamefont {M.}~\bibnamefont {Hu}}, \bibinfo
		{author} {\bibfnamefont {N.}~\bibnamefont {Li}}, \bibinfo {author}
		{\bibfnamefont {J.}~\bibnamefont {Liu}}, \bibinfo {author} {\bibfnamefont
			{W.}~\bibnamefont {Yao}}, \bibinfo {author} {\bibfnamefont {J.-F.}\
			\bibnamefont {Jia}},\ and\ \bibinfo {author} {\bibfnamefont {J.}~\bibnamefont
			{Liu}},\ }\bibfield  {title} {\bibinfo {title} {Multifunctional
			antiferromagnetic materials with giant piezomagnetism and noncollinear spin
			current},\ }\href@noop {} {\bibfield  {journal} {\bibinfo  {journal} {Nat.
				Commun.}\ }\textbf {\bibinfo {volume} {12}},\ \bibinfo {pages} {2846}
		(\bibinfo {year} {2021})}\BibitemShut {NoStop}%
	\bibitem [{\citenamefont {Gonz{\'a}lez-Hern{\'a}ndez}\ \emph
		{et~al.}(2021)\citenamefont {Gonz{\'a}lez-Hern{\'a}ndez}, \citenamefont
		{{\v{S}}mejkal}, \citenamefont {V{\`y}born{\`y}}, \citenamefont {Yahagi},
		\citenamefont {Sinova}, \citenamefont {Jungwirth},\ and\ \citenamefont
		{{\v{Z}}elezn{\`y}}}]{gonzalez2021efficient}%
	\BibitemOpen
	\bibfield  {author} {\bibinfo {author} {\bibfnamefont {R.}~\bibnamefont
			{Gonz{\'a}lez-Hern{\'a}ndez}}, \bibinfo {author} {\bibfnamefont
			{L.}~\bibnamefont {{\v{S}}mejkal}}, \bibinfo {author} {\bibfnamefont
			{K.}~\bibnamefont {V{\`y}born{\`y}}}, \bibinfo {author} {\bibfnamefont
			{Y.}~\bibnamefont {Yahagi}}, \bibinfo {author} {\bibfnamefont
			{J.}~\bibnamefont {Sinova}}, \bibinfo {author} {\bibfnamefont
			{T.}~\bibnamefont {Jungwirth}},\ and\ \bibinfo {author} {\bibfnamefont
			{J.}~\bibnamefont {{\v{Z}}elezn{\`y}}},\ }\bibfield  {title} {\bibinfo
		{title} {Efficient electrical spin splitter based on nonrelativistic
			collinear antiferromagnetism},\ }\href@noop {} {\bibfield  {journal}
		{\bibinfo  {journal} {Phys. Rev. Lett.}\ }\textbf {\bibinfo {volume} {126}},\
		\bibinfo {pages} {127701} (\bibinfo {year} {2021})}\BibitemShut {NoStop}%
	\bibitem [{\citenamefont {Bose}\ \emph {et~al.}(2022)\citenamefont {Bose},
		\citenamefont {Schreiber}, \citenamefont {Jain}, \citenamefont {Shao},
		\citenamefont {Nair}, \citenamefont {Sun}, \citenamefont {Zhang},
		\citenamefont {Muller}, \citenamefont {Tsymbal}, \citenamefont {Schlom} \emph
		{et~al.}}]{Bose22NE}%
	\BibitemOpen
	\bibfield  {author} {\bibinfo {author} {\bibfnamefont {A.}~\bibnamefont
			{Bose}}, \bibinfo {author} {\bibfnamefont {N.~J.}\ \bibnamefont {Schreiber}},
		\bibinfo {author} {\bibfnamefont {R.}~\bibnamefont {Jain}}, \bibinfo {author}
		{\bibfnamefont {D.-F.}\ \bibnamefont {Shao}}, \bibinfo {author}
		{\bibfnamefont {H.~P.}\ \bibnamefont {Nair}}, \bibinfo {author}
		{\bibfnamefont {J.}~\bibnamefont {Sun}}, \bibinfo {author} {\bibfnamefont
			{X.~S.}\ \bibnamefont {Zhang}}, \bibinfo {author} {\bibfnamefont {D.~A.}\
			\bibnamefont {Muller}}, \bibinfo {author} {\bibfnamefont {E.~Y.}\
			\bibnamefont {Tsymbal}}, \bibinfo {author} {\bibfnamefont {D.~G.}\
			\bibnamefont {Schlom}}, \emph {et~al.},\ }\bibfield  {title} {\bibinfo
		{title} {{Tilted spin current generated by the collinear antiferromagnet
				ruthenium dioxide}},\ }\href@noop {} {\bibfield  {journal} {\bibinfo
			{journal} {Nat. Electron.}\ }\textbf {\bibinfo {volume} {5}},\ \bibinfo
		{pages} {267} (\bibinfo {year} {2022})}\BibitemShut {NoStop}%
	\bibitem [{\citenamefont {Shao}\ \emph {et~al.}(2021)\citenamefont {Shao},
		\citenamefont {Zhang}, \citenamefont {Li}, \citenamefont {Eom},\ and\
		\citenamefont {Tsymbal}}]{shao2021spin}%
	\BibitemOpen
	\bibfield  {author} {\bibinfo {author} {\bibfnamefont {D.-F.}\ \bibnamefont
			{Shao}}, \bibinfo {author} {\bibfnamefont {S.-H.}\ \bibnamefont {Zhang}},
		\bibinfo {author} {\bibfnamefont {M.}~\bibnamefont {Li}}, \bibinfo {author}
		{\bibfnamefont {C.-B.}\ \bibnamefont {Eom}},\ and\ \bibinfo {author}
		{\bibfnamefont {E.~Y.}\ \bibnamefont {Tsymbal}},\ }\bibfield  {title}
	{\bibinfo {title} {Spin-neutral currents for spintronics},\ }\href@noop {}
	{\bibfield  {journal} {\bibinfo  {journal} {Nat. Commun.}\ }\textbf {\bibinfo
			{volume} {12}},\ \bibinfo {pages} {7061} (\bibinfo {year}
		{2021})}\BibitemShut {NoStop}%
	\bibitem [{\citenamefont {{\v{S}}mejkal}\ \emph
		{et~al.}(2022{\natexlab{c}})\citenamefont {{\v{S}}mejkal}, \citenamefont
		{Hellenes}, \citenamefont {Gonz{\'a}lez-Hern{\'a}ndez}, \citenamefont
		{Sinova},\ and\ \citenamefont {Jungwirth}}]{vsmejkal2022giant}%
	\BibitemOpen
	\bibfield  {author} {\bibinfo {author} {\bibfnamefont {L.}~\bibnamefont
			{{\v{S}}mejkal}}, \bibinfo {author} {\bibfnamefont {A.~B.}\ \bibnamefont
			{Hellenes}}, \bibinfo {author} {\bibfnamefont {R.}~\bibnamefont
			{Gonz{\'a}lez-Hern{\'a}ndez}}, \bibinfo {author} {\bibfnamefont
			{J.}~\bibnamefont {Sinova}},\ and\ \bibinfo {author} {\bibfnamefont
			{T.}~\bibnamefont {Jungwirth}},\ }\bibfield  {title} {\bibinfo {title} {Giant
			and tunneling magnetoresistance in unconventional collinear antiferromagnets
			with nonrelativistic spin-momentum coupling},\ }\href@noop {} {\bibfield
		{journal} {\bibinfo  {journal} {Phys. Rev. X}\ }\textbf {\bibinfo {volume}
			{12}},\ \bibinfo {pages} {011028} (\bibinfo {year}
		{2022}{\natexlab{c}})}\BibitemShut {NoStop}%
	\bibitem [{\citenamefont {Cui}\ \emph {et~al.}(2023)\citenamefont {Cui},
		\citenamefont {Zeng}, \citenamefont {Cui}, \citenamefont {Yu},\ and\
		\citenamefont {Yang}}]{cui2023efficient}%
	\BibitemOpen
	\bibfield  {author} {\bibinfo {author} {\bibfnamefont {Q.}~\bibnamefont
			{Cui}}, \bibinfo {author} {\bibfnamefont {B.}~\bibnamefont {Zeng}}, \bibinfo
		{author} {\bibfnamefont {P.}~\bibnamefont {Cui}}, \bibinfo {author}
		{\bibfnamefont {T.}~\bibnamefont {Yu}},\ and\ \bibinfo {author}
		{\bibfnamefont {H.}~\bibnamefont {Yang}},\ }\bibfield  {title} {\bibinfo
		{title} {Efficient spin seebeck and spin nernst effects of magnons in
			altermagnets},\ }\href@noop {} {\bibfield  {journal} {\bibinfo  {journal}
			{Phys. Rev. B}\ }\textbf {\bibinfo {volume} {108}},\ \bibinfo {pages}
		{L180401} (\bibinfo {year} {2023})}\BibitemShut {NoStop}%
	\bibitem [{\citenamefont {Papaj}(2023)}]{Papaj23PRB}%
	\BibitemOpen
	\bibfield  {author} {\bibinfo {author} {\bibfnamefont {M.}~\bibnamefont
			{Papaj}},\ }\bibfield  {title} {\bibinfo {title} {Andreev reflection at the
			altermagnet-superconductor interface},\ }\href@noop {} {\bibfield  {journal}
		{\bibinfo  {journal} {Phys. Rev. B}\ }\textbf {\bibinfo {volume} {108}},\
		\bibinfo {pages} {L060508} (\bibinfo {year} {2023})}\BibitemShut {NoStop}%
	\bibitem [{\citenamefont {Sun}\ \emph {et~al.}(2023)\citenamefont {Sun},
		\citenamefont {Brataas},\ and\ \citenamefont {Linder}}]{sun2023andreev}%
	\BibitemOpen
	\bibfield  {author} {\bibinfo {author} {\bibfnamefont {C.}~\bibnamefont
			{Sun}}, \bibinfo {author} {\bibfnamefont {A.}~\bibnamefont {Brataas}},\ and\
		\bibinfo {author} {\bibfnamefont {J.}~\bibnamefont {Linder}},\ }\bibfield
	{title} {\bibinfo {title} {Andreev reflection in altermagnets},\ }\href@noop
	{} {\bibfield  {journal} {\bibinfo  {journal} {Phys. Rev. B}\ }\textbf
		{\bibinfo {volume} {108}},\ \bibinfo {pages} {054511} (\bibinfo {year}
		{2023})}\BibitemShut {NoStop}%
	\bibitem [{\citenamefont {Zhang}\ \emph {et~al.}(2024)\citenamefont {Zhang},
		\citenamefont {Hu},\ and\ \citenamefont {Neupert}}]{Zhang2024Finite}%
	\BibitemOpen
	\bibfield  {author} {\bibinfo {author} {\bibfnamefont {S.-B.}\ \bibnamefont
			{Zhang}}, \bibinfo {author} {\bibfnamefont {L.-H.}\ \bibnamefont {Hu}},\ and\
		\bibinfo {author} {\bibfnamefont {T.}~\bibnamefont {Neupert}},\ }\bibfield
	{title} {\bibinfo {title} {Finite-momentum cooper pairing in proximitized
			altermagnets},\ }\href@noop {} {\bibfield  {journal} {\bibinfo  {journal}
			{Nat. Commun.}\ }\textbf {\bibinfo {volume} {15}},\ \bibinfo {pages} {1801}
		(\bibinfo {year} {2024})}\BibitemShut {NoStop}%
	\bibitem [{\citenamefont {Hong}\ \emph {et~al.}(2025)\citenamefont {Hong},
		\citenamefont {Park},\ and\ \citenamefont {Kim}}]{hong2024unconventional}%
	\BibitemOpen
	\bibfield  {author} {\bibinfo {author} {\bibfnamefont {S.}~\bibnamefont
			{Hong}}, \bibinfo {author} {\bibfnamefont {M.~J.}\ \bibnamefont {Park}},\
		and\ \bibinfo {author} {\bibfnamefont {K.-M.}\ \bibnamefont {Kim}},\
	}\bibfield  {title} {\bibinfo {title} {Unconventional $p$-wave and
			finite-momentum superconductivity induced by altermagnetism through the
			formation of bogoliubov fermi surface},\ }\href@noop {} {\bibfield  {journal}
		{\bibinfo  {journal} {Phys. Rev. B}\ }\textbf {\bibinfo {volume} {111}},\
		\bibinfo {pages} {054501} (\bibinfo {year} {2025})}\BibitemShut {NoStop}%
	\bibitem [{\citenamefont {Sim}\ and\ \citenamefont
		{Knolle}(2025)}]{sim2025pair}%
	\BibitemOpen
	\bibfield  {author} {\bibinfo {author} {\bibfnamefont {G.}~\bibnamefont
			{Sim}}\ and\ \bibinfo {author} {\bibfnamefont {J.}~\bibnamefont {Knolle}},\
	}\bibfield  {title} {\bibinfo {title} {Pair density waves and supercurrent
			diode effect in altermagnets},\ }\href@noop {} {\bibfield  {journal}
		{\bibinfo  {journal} {Phys. Rev. B}\ }\textbf {\bibinfo {volume} {112}},\
		\bibinfo {pages} {L020502} (\bibinfo {year} {2025})}\BibitemShut {NoStop}%
	\bibitem [{\citenamefont {Chakraborty}\ and\ \citenamefont
		{Black-Schaffer}(2024)}]{PhysRevB.110.L060508}%
	\BibitemOpen
	\bibfield  {author} {\bibinfo {author} {\bibfnamefont {D.}~\bibnamefont
			{Chakraborty}}\ and\ \bibinfo {author} {\bibfnamefont {A.~M.}\ \bibnamefont
			{Black-Schaffer}},\ }\bibfield  {title} {\bibinfo {title} {Zero-field
			finite-momentum and field-induced superconductivity in altermagnets},\
	}\href@noop {} {\bibfield  {journal} {\bibinfo  {journal} {Phys. Rev. B}\
		}\textbf {\bibinfo {volume} {110}},\ \bibinfo {pages} {L060508} (\bibinfo
		{year} {2024})}\BibitemShut {NoStop}%
	\bibitem [{\citenamefont {Li}\ and\ \citenamefont
		{Liu}(2023)}]{li2023majorana}%
	\BibitemOpen
	\bibfield  {author} {\bibinfo {author} {\bibfnamefont {Y.-X.}\ \bibnamefont
			{Li}}\ and\ \bibinfo {author} {\bibfnamefont {C.-C.}\ \bibnamefont {Liu}},\
	}\bibfield  {title} {\bibinfo {title} {Majorana corner modes and tunable
			patterns in an altermagnet heterostructure},\ }\href@noop {} {\bibfield
		{journal} {\bibinfo  {journal} {Phys. Rev. B}\ }\textbf {\bibinfo {volume}
			{108}},\ \bibinfo {pages} {205410} (\bibinfo {year} {2023})}\BibitemShut
	{NoStop}%
	\bibitem [{\citenamefont {Ghorashi}\ \emph {et~al.}(2024)\citenamefont
		{Ghorashi}, \citenamefont {Hughes},\ and\ \citenamefont
		{Cano}}]{PhysRevLett.133.106601}%
	\BibitemOpen
	\bibfield  {author} {\bibinfo {author} {\bibfnamefont {S.~A.~A.}\
			\bibnamefont {Ghorashi}}, \bibinfo {author} {\bibfnamefont {T.~L.}\
			\bibnamefont {Hughes}},\ and\ \bibinfo {author} {\bibfnamefont
			{J.}~\bibnamefont {Cano}},\ }\bibfield  {title} {\bibinfo {title}
		{Altermagnetic routes to majorana modes in zero net magnetization},\
	}\href@noop {} {\bibfield  {journal} {\bibinfo  {journal} {Phys. Rev. Lett.}\
		}\textbf {\bibinfo {volume} {133}},\ \bibinfo {pages} {106601} (\bibinfo
		{year} {2024})}\BibitemShut {NoStop}%
	\bibitem [{\citenamefont {Li}\ \emph {et~al.}(2024)\citenamefont {Li},
		\citenamefont {Liu},\ and\ \citenamefont {Liu}}]{PhysRevB.109.L201109}%
	\BibitemOpen
	\bibfield  {author} {\bibinfo {author} {\bibfnamefont {Y.-X.}\ \bibnamefont
			{Li}}, \bibinfo {author} {\bibfnamefont {Y.}~\bibnamefont {Liu}},\ and\
		\bibinfo {author} {\bibfnamefont {C.-C.}\ \bibnamefont {Liu}},\ }\bibfield
	{title} {\bibinfo {title} {Creation and manipulation of higher-order
			topological states by altermagnets},\ }\href@noop {} {\bibfield  {journal}
		{\bibinfo  {journal} {Phys. Rev. B}\ }\textbf {\bibinfo {volume} {109}},\
		\bibinfo {pages} {L201109} (\bibinfo {year} {2024})}\BibitemShut {NoStop}%
	\bibitem [{\citenamefont {Zhu}\ \emph {et~al.}(2023)\citenamefont {Zhu},
		\citenamefont {Zhuang}, \citenamefont {Wu},\ and\ \citenamefont
		{Yan}}]{zhu2023topological}%
	\BibitemOpen
	\bibfield  {author} {\bibinfo {author} {\bibfnamefont {D.}~\bibnamefont
			{Zhu}}, \bibinfo {author} {\bibfnamefont {Z.-Y.}\ \bibnamefont {Zhuang}},
		\bibinfo {author} {\bibfnamefont {Z.}~\bibnamefont {Wu}},\ and\ \bibinfo
		{author} {\bibfnamefont {Z.}~\bibnamefont {Yan}},\ }\bibfield  {title}
	{\bibinfo {title} {Topological superconductivity in two-dimensional
			altermagnetic metals},\ }\href@noop {} {\bibfield  {journal} {\bibinfo
			{journal} {Phys. Rev. B}\ }\textbf {\bibinfo {volume} {108}},\ \bibinfo
		{pages} {184505} (\bibinfo {year} {2023})}\BibitemShut {NoStop}%
	\bibitem [{\citenamefont {Gu}\ \emph {et~al.}(2025)\citenamefont {Gu},
		\citenamefont {Liu}, \citenamefont {Zhu}, \citenamefont {Yananose},
		\citenamefont {Chen}, \citenamefont {Hu}, \citenamefont {Stroppa},\ and\
		\citenamefont {Liu}}]{gu2025ferroelectric}%
	\BibitemOpen
	\bibfield  {author} {\bibinfo {author} {\bibfnamefont {M.}~\bibnamefont
			{Gu}}, \bibinfo {author} {\bibfnamefont {Y.}~\bibnamefont {Liu}}, \bibinfo
		{author} {\bibfnamefont {H.}~\bibnamefont {Zhu}}, \bibinfo {author}
		{\bibfnamefont {K.}~\bibnamefont {Yananose}}, \bibinfo {author}
		{\bibfnamefont {X.}~\bibnamefont {Chen}}, \bibinfo {author} {\bibfnamefont
			{Y.}~\bibnamefont {Hu}}, \bibinfo {author} {\bibfnamefont {A.}~\bibnamefont
			{Stroppa}},\ and\ \bibinfo {author} {\bibfnamefont {Q.}~\bibnamefont {Liu}},\
	}\bibfield  {title} {\bibinfo {title} {Ferroelectric switchable
			altermagnetism},\ }\href@noop {} {\bibfield  {journal} {\bibinfo  {journal}
			{Phys. Rev. Lett.}\ }\textbf {\bibinfo {volume} {134}},\ \bibinfo {pages}
		{106802} (\bibinfo {year} {2025})}\BibitemShut {NoStop}%
	\bibitem [{\citenamefont {Duan}\ \emph {et~al.}(2025)\citenamefont {Duan},
		\citenamefont {Zhang}, \citenamefont {Zhu}, \citenamefont {Liu},
		\citenamefont {Zhang}, \citenamefont {{\v{Z}}uti{\'c}},\ and\ \citenamefont
		{Zhou}}]{duan2025antiferroelectric}%
	\BibitemOpen
	\bibfield  {author} {\bibinfo {author} {\bibfnamefont {X.}~\bibnamefont
			{Duan}}, \bibinfo {author} {\bibfnamefont {J.}~\bibnamefont {Zhang}},
		\bibinfo {author} {\bibfnamefont {Z.}~\bibnamefont {Zhu}}, \bibinfo {author}
		{\bibfnamefont {Y.}~\bibnamefont {Liu}}, \bibinfo {author} {\bibfnamefont
			{Z.}~\bibnamefont {Zhang}}, \bibinfo {author} {\bibfnamefont
			{I.}~\bibnamefont {{\v{Z}}uti{\'c}}},\ and\ \bibinfo {author} {\bibfnamefont
			{T.}~\bibnamefont {Zhou}},\ }\bibfield  {title} {\bibinfo {title}
		{Antiferroelectric altermagnets: Antiferroelectricity alters magnets},\
	}\href@noop {} {\bibfield  {journal} {\bibinfo  {journal} {Phys. Rev. Lett.}\
		}\textbf {\bibinfo {volume} {134}},\ \bibinfo {pages} {106801} (\bibinfo
		{year} {2025})}\BibitemShut {NoStop}%
	\bibitem [{\citenamefont {{\v{S}}mejkal}()}]{vsmejkal2024altermagnetic}%
	\BibitemOpen
	\bibfield  {author} {\bibinfo {author} {\bibfnamefont {L.}~\bibnamefont
			{{\v{S}}mejkal}},\ }\bibfield  {title} {\bibinfo {title} {Altermagnetic
			multiferroics and altermagnetoelectric effect},\ }\href@noop {} {\bibinfo
		{journal} {arXiv:2411.19928}\ }\BibitemShut {NoStop}%
	\bibitem [{\citenamefont {Urru}\ \emph {et~al.}(2025)\citenamefont {Urru},
		\citenamefont {Seleznev}, \citenamefont {Teng}, \citenamefont {Park},
		\citenamefont {Reyes-Lillo},\ and\ \citenamefont {Rabe}}]{urru2025g}%
	\BibitemOpen
	\bibfield  {journal} {  }\bibfield  {author} {\bibinfo {author} {\bibfnamefont
			{A.}~\bibnamefont {Urru}}, \bibinfo {author} {\bibfnamefont {D.}~\bibnamefont
			{Seleznev}}, \bibinfo {author} {\bibfnamefont {Y.}~\bibnamefont {Teng}},
		\bibinfo {author} {\bibfnamefont {S.~Y.}\ \bibnamefont {Park}}, \bibinfo
		{author} {\bibfnamefont {S.~E.}\ \bibnamefont {Reyes-Lillo}},\ and\ \bibinfo
		{author} {\bibfnamefont {K.~M.}\ \bibnamefont {Rabe}},\ }\bibfield  {title}
	{\bibinfo {title} {G-type antiferromagnetic bifeo 3 is a multiferroic g-wave
			altermagnet},\ }\href@noop {} {\bibfield  {journal} {\bibinfo  {journal}
			{Phys. Rev. B}\ }\textbf {\bibinfo {volume} {112}},\ \bibinfo {pages}
		{104411} (\bibinfo {year} {2025})}\BibitemShut {NoStop}%
	\bibitem [{\citenamefont {Berlijn}\ \emph {et~al.}(2017)\citenamefont
		{Berlijn}, \citenamefont {Snijders}, \citenamefont {Delaire}, \citenamefont
		{Zhou}, \citenamefont {Maier}, \citenamefont {Cao}, \citenamefont {Chi},
		\citenamefont {Matsuda}, \citenamefont {Wang}, \citenamefont {Koehler} \emph
		{et~al.}}]{berlijn2017itinerant}%
	\BibitemOpen
	\bibfield  {author} {\bibinfo {author} {\bibfnamefont {T.}~\bibnamefont
			{Berlijn}}, \bibinfo {author} {\bibfnamefont {P.~C.}\ \bibnamefont
			{Snijders}}, \bibinfo {author} {\bibfnamefont {O.}~\bibnamefont {Delaire}},
		\bibinfo {author} {\bibfnamefont {H.-D.}\ \bibnamefont {Zhou}}, \bibinfo
		{author} {\bibfnamefont {T.~A.}\ \bibnamefont {Maier}}, \bibinfo {author}
		{\bibfnamefont {H.-B.}\ \bibnamefont {Cao}}, \bibinfo {author} {\bibfnamefont
			{S.-X.}\ \bibnamefont {Chi}}, \bibinfo {author} {\bibfnamefont
			{M.}~\bibnamefont {Matsuda}}, \bibinfo {author} {\bibfnamefont
			{Y.}~\bibnamefont {Wang}}, \bibinfo {author} {\bibfnamefont {M.~R.}\
			\bibnamefont {Koehler}}, \emph {et~al.},\ }\bibfield  {title} {\bibinfo
		{title} {Itinerant antiferromagnetism in ruo 2},\ }\href@noop {} {\bibfield
		{journal} {\bibinfo  {journal} {Phys. Rev. Lett.}\ }\textbf {\bibinfo
			{volume} {118}},\ \bibinfo {pages} {077201} (\bibinfo {year}
		{2017})}\BibitemShut {NoStop}%
	\bibitem [{\citenamefont {Zhu}\ \emph {et~al.}(2019)\citenamefont {Zhu},
		\citenamefont {Strempfer}, \citenamefont {Rao}, \citenamefont {Occhialini},
		\citenamefont {Pelliciari}, \citenamefont {Choi}, \citenamefont {Kawaguchi},
		\citenamefont {You}, \citenamefont {Mitchell}, \citenamefont {Shao-Horn}
		\emph {et~al.}}]{zhu2019anomalous}%
	\BibitemOpen
	\bibfield  {author} {\bibinfo {author} {\bibfnamefont {Z.}~\bibnamefont
			{Zhu}}, \bibinfo {author} {\bibfnamefont {J.}~\bibnamefont {Strempfer}},
		\bibinfo {author} {\bibfnamefont {R.}~\bibnamefont {Rao}}, \bibinfo {author}
		{\bibfnamefont {C.}~\bibnamefont {Occhialini}}, \bibinfo {author}
		{\bibfnamefont {J.}~\bibnamefont {Pelliciari}}, \bibinfo {author}
		{\bibfnamefont {Y.}~\bibnamefont {Choi}}, \bibinfo {author} {\bibfnamefont
			{T.}~\bibnamefont {Kawaguchi}}, \bibinfo {author} {\bibfnamefont
			{H.}~\bibnamefont {You}}, \bibinfo {author} {\bibfnamefont {J.}~\bibnamefont
			{Mitchell}}, \bibinfo {author} {\bibfnamefont {Y.}~\bibnamefont {Shao-Horn}},
		\emph {et~al.},\ }\bibfield  {title} {\bibinfo {title} {Anomalous
			antiferromagnetism in metallic $\mathrm{RuO}_{2}$ determined by resonant
			x-ray scattering},\ }\href@noop {} {\bibfield  {journal} {\bibinfo  {journal}
			{Phys. Rev. Lett.}\ }\textbf {\bibinfo {volume} {122}},\ \bibinfo {pages}
		{017202} (\bibinfo {year} {2019})}\BibitemShut {NoStop}%
	\bibitem [{\citenamefont {Fedchenko}\ \emph {et~al.}(2024)\citenamefont
		{Fedchenko}, \citenamefont {Min{\'a}r}, \citenamefont {Akashdeep},
		\citenamefont {D’Souza}, \citenamefont {Vasilyev}, \citenamefont {Tkach},
		\citenamefont {Odenbreit}, \citenamefont {Nguyen}, \citenamefont
		{Kutnyakhov}, \citenamefont {Wind} \emph
		{et~al.}}]{fedchenko2024observation}%
	\BibitemOpen
	\bibfield  {author} {\bibinfo {author} {\bibfnamefont {O.}~\bibnamefont
			{Fedchenko}}, \bibinfo {author} {\bibfnamefont {J.}~\bibnamefont
			{Min{\'a}r}}, \bibinfo {author} {\bibfnamefont {A.}~\bibnamefont
			{Akashdeep}}, \bibinfo {author} {\bibfnamefont {S.~W.}\ \bibnamefont
			{D’Souza}}, \bibinfo {author} {\bibfnamefont {D.}~\bibnamefont {Vasilyev}},
		\bibinfo {author} {\bibfnamefont {O.}~\bibnamefont {Tkach}}, \bibinfo
		{author} {\bibfnamefont {L.}~\bibnamefont {Odenbreit}}, \bibinfo {author}
		{\bibfnamefont {Q.}~\bibnamefont {Nguyen}}, \bibinfo {author} {\bibfnamefont
			{D.}~\bibnamefont {Kutnyakhov}}, \bibinfo {author} {\bibfnamefont
			{N.}~\bibnamefont {Wind}}, \emph {et~al.},\ }\bibfield  {title} {\bibinfo
		{title} {Observation of time-reversal symmetry breaking in the band structure
			of altermagnetic $\mathrm{RuO}_{2}$},\ }\href@noop {} {\bibfield  {journal}
		{\bibinfo  {journal} {Sci. Adv.}\ }\textbf {\bibinfo {volume} {10}},\
		\bibinfo {pages} {eadj4883} (\bibinfo {year} {2024})}\BibitemShut {NoStop}%
	\bibitem [{\citenamefont {Gonzalez~Betancourt}\ \emph
		{et~al.}(2023)\citenamefont {Gonzalez~Betancourt}, \citenamefont
		{Zub{\'a}{\v{c}}}, \citenamefont {Gonzalez-Hernandez}, \citenamefont
		{Geishendorf}, \citenamefont {{\v{S}}ob{\'a}{\v{n}}}, \citenamefont
		{Springholz}, \citenamefont {Olejn{\'\i}k}, \citenamefont {{\v{S}}mejkal},
		\citenamefont {Sinova}, \citenamefont {Jungwirth} \emph
		{et~al.}}]{gonzalez2023spontaneous}%
	\BibitemOpen
	\bibfield  {author} {\bibinfo {author} {\bibfnamefont {R.}~\bibnamefont
			{Gonzalez~Betancourt}}, \bibinfo {author} {\bibfnamefont {J.}~\bibnamefont
			{Zub{\'a}{\v{c}}}}, \bibinfo {author} {\bibfnamefont {R.}~\bibnamefont
			{Gonzalez-Hernandez}}, \bibinfo {author} {\bibfnamefont {K.}~\bibnamefont
			{Geishendorf}}, \bibinfo {author} {\bibfnamefont {Z.}~\bibnamefont
			{{\v{S}}ob{\'a}{\v{n}}}}, \bibinfo {author} {\bibfnamefont {G.}~\bibnamefont
			{Springholz}}, \bibinfo {author} {\bibfnamefont {K.}~\bibnamefont
			{Olejn{\'\i}k}}, \bibinfo {author} {\bibfnamefont {L.}~\bibnamefont
			{{\v{S}}mejkal}}, \bibinfo {author} {\bibfnamefont {J.}~\bibnamefont
			{Sinova}}, \bibinfo {author} {\bibfnamefont {T.}~\bibnamefont {Jungwirth}},
		\emph {et~al.},\ }\bibfield  {title} {\bibinfo {title} {Spontaneous anomalous
			hall effect arising from an unconventional compensated magnetic phase in a
			semiconductor},\ }\href@noop {} {\bibfield  {journal} {\bibinfo  {journal}
			{Phys. Rev. Lett.}\ }\textbf {\bibinfo {volume} {130}},\ \bibinfo {pages}
		{036702} (\bibinfo {year} {2023})}\BibitemShut {NoStop}%
	\bibitem [{\citenamefont {Krempask{\`y}}\ \emph {et~al.}(2024)\citenamefont
		{Krempask{\`y}}, \citenamefont {{\v{S}}mejkal}, \citenamefont {D’souza},
		\citenamefont {Hajlaoui}, \citenamefont {Springholz}, \citenamefont
		{Uhl{\'\i}{\v{r}}ov{\'a}}, \citenamefont {Alarab}, \citenamefont
		{Constantinou}, \citenamefont {Strocov}, \citenamefont {Usanov} \emph
		{et~al.}}]{krempasky2024altermagnetic}%
	\BibitemOpen
	\bibfield  {author} {\bibinfo {author} {\bibfnamefont {J.}~\bibnamefont
			{Krempask{\`y}}}, \bibinfo {author} {\bibfnamefont {L.}~\bibnamefont
			{{\v{S}}mejkal}}, \bibinfo {author} {\bibfnamefont {S.}~\bibnamefont
			{D’souza}}, \bibinfo {author} {\bibfnamefont {M.}~\bibnamefont {Hajlaoui}},
		\bibinfo {author} {\bibfnamefont {G.}~\bibnamefont {Springholz}}, \bibinfo
		{author} {\bibfnamefont {K.}~\bibnamefont {Uhl{\'\i}{\v{r}}ov{\'a}}},
		\bibinfo {author} {\bibfnamefont {F.}~\bibnamefont {Alarab}}, \bibinfo
		{author} {\bibfnamefont {P.}~\bibnamefont {Constantinou}}, \bibinfo {author}
		{\bibfnamefont {V.}~\bibnamefont {Strocov}}, \bibinfo {author} {\bibfnamefont
			{D.}~\bibnamefont {Usanov}}, \emph {et~al.},\ }\bibfield  {title} {\bibinfo
		{title} {Altermagnetic lifting of kramers spin degeneracy},\ }\href@noop {}
	{\bibfield  {journal} {\bibinfo  {journal} {Nature}\ }\textbf {\bibinfo
			{volume} {626}},\ \bibinfo {pages} {517} (\bibinfo {year}
		{2024})}\BibitemShut {NoStop}%
	\bibitem [{\citenamefont {Zhu}\ \emph {et~al.}(2024)\citenamefont {Zhu},
		\citenamefont {Chen}, \citenamefont {Liu}, \citenamefont {Liu}, \citenamefont
		{Liu}, \citenamefont {Zha}, \citenamefont {Qu}, \citenamefont {Hong},
		\citenamefont {Li}, \citenamefont {Jiang} \emph
		{et~al.}}]{zhu2024observation}%
	\BibitemOpen
	\bibfield  {author} {\bibinfo {author} {\bibfnamefont {Y.-P.}\ \bibnamefont
			{Zhu}}, \bibinfo {author} {\bibfnamefont {X.}~\bibnamefont {Chen}}, \bibinfo
		{author} {\bibfnamefont {X.-R.}\ \bibnamefont {Liu}}, \bibinfo {author}
		{\bibfnamefont {Y.}~\bibnamefont {Liu}}, \bibinfo {author} {\bibfnamefont
			{P.}~\bibnamefont {Liu}}, \bibinfo {author} {\bibfnamefont {H.}~\bibnamefont
			{Zha}}, \bibinfo {author} {\bibfnamefont {G.}~\bibnamefont {Qu}}, \bibinfo
		{author} {\bibfnamefont {C.}~\bibnamefont {Hong}}, \bibinfo {author}
		{\bibfnamefont {J.}~\bibnamefont {Li}}, \bibinfo {author} {\bibfnamefont
			{Z.}~\bibnamefont {Jiang}}, \emph {et~al.},\ }\bibfield  {title} {\bibinfo
		{title} {Observation of plaid-like spin splitting in a noncoplanar
			antiferromagnet},\ }\href@noop {} {\bibfield  {journal} {\bibinfo  {journal}
			{Nature}\ }\textbf {\bibinfo {volume} {626}},\ \bibinfo {pages} {523}
		(\bibinfo {year} {2024})}\BibitemShut {NoStop}%
	\bibitem [{\citenamefont {Li}\ \emph {et~al.}(2025)\citenamefont {Li},
		\citenamefont {Hu}, \citenamefont {Li}, \citenamefont {Wang}, \citenamefont
		{Chen}, \citenamefont {Thiagarajan}, \citenamefont {Leandersson},
		\citenamefont {Polley}, \citenamefont {Kim}, \citenamefont {Liu} \emph
		{et~al.}}]{li2024topological}%
	\BibitemOpen
	\bibfield  {author} {\bibinfo {author} {\bibfnamefont {C.}~\bibnamefont
			{Li}}, \bibinfo {author} {\bibfnamefont {M.}~\bibnamefont {Hu}}, \bibinfo
		{author} {\bibfnamefont {Z.}~\bibnamefont {Li}}, \bibinfo {author}
		{\bibfnamefont {Y.}~\bibnamefont {Wang}}, \bibinfo {author} {\bibfnamefont
			{W.}~\bibnamefont {Chen}}, \bibinfo {author} {\bibfnamefont {B.}~\bibnamefont
			{Thiagarajan}}, \bibinfo {author} {\bibfnamefont {M.}~\bibnamefont
			{Leandersson}}, \bibinfo {author} {\bibfnamefont {C.}~\bibnamefont {Polley}},
		\bibinfo {author} {\bibfnamefont {T.}~\bibnamefont {Kim}}, \bibinfo {author}
		{\bibfnamefont {H.}~\bibnamefont {Liu}}, \emph {et~al.},\ }\bibfield  {title}
	{\bibinfo {title} {Topological $\mathrm{Weyl}$ altermagnetism in
			$\mathrm{CrSb}$},\ }\href@noop {} {\bibfield  {journal} {\bibinfo  {journal}
			{Commun. Phys.}\ }\textbf {\bibinfo {volume} {8}},\ \bibinfo {pages} {311}
		(\bibinfo {year} {2025})}\BibitemShut {NoStop}%
	\bibitem [{\citenamefont {Lu}\ \emph {et~al.}(2025)\citenamefont {Lu},
		\citenamefont {Feng}, \citenamefont {Wang}, \citenamefont {Chen},
		\citenamefont {Lin}, \citenamefont {Liang}, \citenamefont {Liu},
		\citenamefont {Feng}, \citenamefont {Yamagami}, \citenamefont {Liu} \emph
		{et~al.}}]{lu2025signature}%
	\BibitemOpen
	\bibfield  {author} {\bibinfo {author} {\bibfnamefont {W.}~\bibnamefont
			{Lu}}, \bibinfo {author} {\bibfnamefont {S.}~\bibnamefont {Feng}}, \bibinfo
		{author} {\bibfnamefont {Y.}~\bibnamefont {Wang}}, \bibinfo {author}
		{\bibfnamefont {D.}~\bibnamefont {Chen}}, \bibinfo {author} {\bibfnamefont
			{Z.}~\bibnamefont {Lin}}, \bibinfo {author} {\bibfnamefont {X.}~\bibnamefont
			{Liang}}, \bibinfo {author} {\bibfnamefont {S.}~\bibnamefont {Liu}}, \bibinfo
		{author} {\bibfnamefont {W.}~\bibnamefont {Feng}}, \bibinfo {author}
		{\bibfnamefont {K.}~\bibnamefont {Yamagami}}, \bibinfo {author}
		{\bibfnamefont {J.}~\bibnamefont {Liu}}, \emph {et~al.},\ }\bibfield  {title}
	{\bibinfo {title} {Signature of topological surface bands in altermagnetic
			$\mathrm{Weyl}$ semimetal $\mathrm{CrSb}$},\ }\href@noop {} {\bibfield
		{journal} {\bibinfo  {journal} {Nano Lett.}\ }\textbf {\bibinfo {volume}
			{25}},\ \bibinfo {pages} {7343} (\bibinfo {year} {2025})}\BibitemShut
	{NoStop}%
	\bibitem [{\citenamefont {Ding}\ \emph {et~al.}(2024)\citenamefont {Ding},
		\citenamefont {Jiang}, \citenamefont {Chen}, \citenamefont {Tao},
		\citenamefont {Liu}, \citenamefont {Li}, \citenamefont {Liu}, \citenamefont
		{Sun}, \citenamefont {Cheng}, \citenamefont {Liu} \emph
		{et~al.}}]{ding2024large}%
	\BibitemOpen
	\bibfield  {author} {\bibinfo {author} {\bibfnamefont {J.}~\bibnamefont
			{Ding}}, \bibinfo {author} {\bibfnamefont {Z.}~\bibnamefont {Jiang}},
		\bibinfo {author} {\bibfnamefont {X.}~\bibnamefont {Chen}}, \bibinfo {author}
		{\bibfnamefont {Z.}~\bibnamefont {Tao}}, \bibinfo {author} {\bibfnamefont
			{Z.}~\bibnamefont {Liu}}, \bibinfo {author} {\bibfnamefont {T.}~\bibnamefont
			{Li}}, \bibinfo {author} {\bibfnamefont {J.}~\bibnamefont {Liu}}, \bibinfo
		{author} {\bibfnamefont {J.}~\bibnamefont {Sun}}, \bibinfo {author}
		{\bibfnamefont {J.}~\bibnamefont {Cheng}}, \bibinfo {author} {\bibfnamefont
			{J.}~\bibnamefont {Liu}}, \emph {et~al.},\ }\bibfield  {title} {\bibinfo
		{title} {Large band splitting in g-wave altermagnet $\mathrm{CrSb}$},\
	}\href@noop {} {\bibfield  {journal} {\bibinfo  {journal} {Phys. Rev. Lett.}\
		}\textbf {\bibinfo {volume} {133}},\ \bibinfo {pages} {206401} (\bibinfo
		{year} {2024})}\BibitemShut {NoStop}%
	\bibitem [{\citenamefont {Zhou}\ \emph {et~al.}(2025)\citenamefont {Zhou},
		\citenamefont {Cheng}, \citenamefont {Hu}, \citenamefont {Chu}, \citenamefont
		{Bai}, \citenamefont {Han}, \citenamefont {Liu}, \citenamefont {Pan},\ and\
		\citenamefont {Song}}]{zhou2025manipulation}%
	\BibitemOpen
	\bibfield  {author} {\bibinfo {author} {\bibfnamefont {Z.}~\bibnamefont
			{Zhou}}, \bibinfo {author} {\bibfnamefont {X.}~\bibnamefont {Cheng}},
		\bibinfo {author} {\bibfnamefont {M.}~\bibnamefont {Hu}}, \bibinfo {author}
		{\bibfnamefont {R.}~\bibnamefont {Chu}}, \bibinfo {author} {\bibfnamefont
			{H.}~\bibnamefont {Bai}}, \bibinfo {author} {\bibfnamefont {L.}~\bibnamefont
			{Han}}, \bibinfo {author} {\bibfnamefont {J.}~\bibnamefont {Liu}}, \bibinfo
		{author} {\bibfnamefont {F.}~\bibnamefont {Pan}},\ and\ \bibinfo {author}
		{\bibfnamefont {C.}~\bibnamefont {Song}},\ }\bibfield  {title} {\bibinfo
		{title} {Manipulation of the altermagnetic order in $\mathrm{CrSb}$ via
			crystal symmetry},\ }\href@noop {} {\bibfield  {journal} {\bibinfo  {journal}
			{Nature}\ }\textbf {\bibinfo {volume} {638}},\ \bibinfo {pages} {645}
		(\bibinfo {year} {2025})}\BibitemShut {NoStop}%
	\bibitem [{\citenamefont {Yang}\ \emph {et~al.}(2025)\citenamefont {Yang},
		\citenamefont {Li}, \citenamefont {Yang}, \citenamefont {Li}, \citenamefont
		{Zheng}, \citenamefont {Zhu}, \citenamefont {Pan}, \citenamefont {Xu},
		\citenamefont {Cao}, \citenamefont {Zhao} \emph {et~al.}}]{yang2025three}%
	\BibitemOpen
	\bibfield  {author} {\bibinfo {author} {\bibfnamefont {G.}~\bibnamefont
			{Yang}}, \bibinfo {author} {\bibfnamefont {Z.}~\bibnamefont {Li}}, \bibinfo
		{author} {\bibfnamefont {S.}~\bibnamefont {Yang}}, \bibinfo {author}
		{\bibfnamefont {J.}~\bibnamefont {Li}}, \bibinfo {author} {\bibfnamefont
			{H.}~\bibnamefont {Zheng}}, \bibinfo {author} {\bibfnamefont
			{W.}~\bibnamefont {Zhu}}, \bibinfo {author} {\bibfnamefont {Z.}~\bibnamefont
			{Pan}}, \bibinfo {author} {\bibfnamefont {Y.}~\bibnamefont {Xu}}, \bibinfo
		{author} {\bibfnamefont {S.}~\bibnamefont {Cao}}, \bibinfo {author}
		{\bibfnamefont {W.}~\bibnamefont {Zhao}}, \emph {et~al.},\ }\bibfield
	{title} {\bibinfo {title} {Three-dimensional mapping of the altermagnetic
			spin splitting in $\mathrm{CrSb}$},\ }\href@noop {} {\bibfield  {journal}
		{\bibinfo  {journal} {Nat. Commun.}\ }\textbf {\bibinfo {volume} {16}},\
		\bibinfo {pages} {1442} (\bibinfo {year} {2025})}\BibitemShut {NoStop}%
	\bibitem [{\citenamefont {Zhang}\ \emph {et~al.}(2025)\citenamefont {Zhang},
		\citenamefont {Cheng}, \citenamefont {Yin}, \citenamefont {Liu},
		\citenamefont {Deng}, \citenamefont {Qiao}, \citenamefont {Shi},
		\citenamefont {Zhang}, \citenamefont {Lin}, \citenamefont {Liu} \emph
		{et~al.}}]{zhang2025crystal}%
	\BibitemOpen
	\bibfield  {author} {\bibinfo {author} {\bibfnamefont {F.}~\bibnamefont
			{Zhang}}, \bibinfo {author} {\bibfnamefont {X.}~\bibnamefont {Cheng}},
		\bibinfo {author} {\bibfnamefont {Z.}~\bibnamefont {Yin}}, \bibinfo {author}
		{\bibfnamefont {C.}~\bibnamefont {Liu}}, \bibinfo {author} {\bibfnamefont
			{L.}~\bibnamefont {Deng}}, \bibinfo {author} {\bibfnamefont {Y.}~\bibnamefont
			{Qiao}}, \bibinfo {author} {\bibfnamefont {Z.}~\bibnamefont {Shi}}, \bibinfo
		{author} {\bibfnamefont {S.}~\bibnamefont {Zhang}}, \bibinfo {author}
		{\bibfnamefont {J.}~\bibnamefont {Lin}}, \bibinfo {author} {\bibfnamefont
			{Z.}~\bibnamefont {Liu}}, \emph {et~al.},\ }\bibfield  {title} {\bibinfo
		{title} {Crystal-symmetry-paired spin--valley locking in a layered
			room-temperature metallic altermagnet candidate},\ }\href@noop {} {\bibfield
		{journal} {\bibinfo  {journal} {Nat. Phys.}\ }\textbf {\bibinfo {volume}
			{21}},\ \bibinfo {pages} {760} (\bibinfo {year} {2025})}\BibitemShut
	{NoStop}%
	\bibitem [{\citenamefont {Jiang}\ \emph {et~al.}(2025)\citenamefont {Jiang},
		\citenamefont {Hu}, \citenamefont {Bai}, \citenamefont {Song}, \citenamefont
		{Mu}, \citenamefont {Qu}, \citenamefont {Li}, \citenamefont {Zhu},
		\citenamefont {Pi}, \citenamefont {Wei} \emph {et~al.}}]{jiang2025metallic}%
	\BibitemOpen
	\bibfield  {author} {\bibinfo {author} {\bibfnamefont {B.}~\bibnamefont
			{Jiang}}, \bibinfo {author} {\bibfnamefont {M.}~\bibnamefont {Hu}}, \bibinfo
		{author} {\bibfnamefont {J.}~\bibnamefont {Bai}}, \bibinfo {author}
		{\bibfnamefont {Z.}~\bibnamefont {Song}}, \bibinfo {author} {\bibfnamefont
			{C.}~\bibnamefont {Mu}}, \bibinfo {author} {\bibfnamefont {G.}~\bibnamefont
			{Qu}}, \bibinfo {author} {\bibfnamefont {W.}~\bibnamefont {Li}}, \bibinfo
		{author} {\bibfnamefont {W.}~\bibnamefont {Zhu}}, \bibinfo {author}
		{\bibfnamefont {H.}~\bibnamefont {Pi}}, \bibinfo {author} {\bibfnamefont
			{Z.}~\bibnamefont {Wei}}, \emph {et~al.},\ }\bibfield  {title} {\bibinfo
		{title} {A metallic room-temperature d-wave altermagnet},\ }\href@noop {}
	{\bibfield  {journal} {\bibinfo  {journal} {Nat. Phys.}\ ,\ \bibinfo {pages}
			{1}} (\bibinfo {year} {2025})}\BibitemShut {NoStop}%
	\bibitem [{\citenamefont {Han}\ \emph {et~al.}(2025)\citenamefont {Han},
		\citenamefont {He}, \citenamefont {Zhang}, \citenamefont {Li}, \citenamefont
		{Yu},\ and\ \citenamefont {Yao}}]{han2025real}%
	\BibitemOpen
	\bibfield  {author} {\bibinfo {author} {\bibfnamefont {Y.}~\bibnamefont
			{Han}}, \bibinfo {author} {\bibfnamefont {T.}~\bibnamefont {He}}, \bibinfo
		{author} {\bibfnamefont {R.-W.}\ \bibnamefont {Zhang}}, \bibinfo {author}
		{\bibfnamefont {Z.}~\bibnamefont {Li}}, \bibinfo {author} {\bibfnamefont
			{Z.-M.}\ \bibnamefont {Yu}},\ and\ \bibinfo {author} {\bibfnamefont
			{Y.}~\bibnamefont {Yao}},\ }\bibfield  {title} {\bibinfo {title} {Real chern
			insulator in monolayer decorated transition metal nitrides},\ }\href@noop {}
	{\bibfield  {journal} {\bibinfo  {journal} {Adv. Funct. Mater.}\ ,\ \bibinfo
			{pages} {2505282}} (\bibinfo {year} {2025})}\BibitemShut {NoStop}%
	\bibitem [{\citenamefont {Wang}\ \emph {et~al.}(2025)\citenamefont {Wang},
		\citenamefont {Yang}, \citenamefont {Yang}, \citenamefont {Lu}, \citenamefont
		{Ho}, \citenamefont {Wang}, \citenamefont {Ang}, \citenamefont {Cheng},\ and\
		\citenamefont {Fang}}]{wang2025pentagonal}%
	\BibitemOpen
	\bibfield  {author} {\bibinfo {author} {\bibfnamefont {J.}~\bibnamefont
			{Wang}}, \bibinfo {author} {\bibfnamefont {X.}~\bibnamefont {Yang}}, \bibinfo
		{author} {\bibfnamefont {Z.}~\bibnamefont {Yang}}, \bibinfo {author}
		{\bibfnamefont {J.}~\bibnamefont {Lu}}, \bibinfo {author} {\bibfnamefont
			{P.}~\bibnamefont {Ho}}, \bibinfo {author} {\bibfnamefont {W.}~\bibnamefont
			{Wang}}, \bibinfo {author} {\bibfnamefont {Y.~S.}\ \bibnamefont {Ang}},
		\bibinfo {author} {\bibfnamefont {Z.}~\bibnamefont {Cheng}},\ and\ \bibinfo
		{author} {\bibfnamefont {S.}~\bibnamefont {Fang}},\ }\bibfield  {title}
	{\bibinfo {title} {Pentagonal $\mathrm{2D}$ altermagnets: Material screening
			and altermagnetic tunneling junction device application},\ }\href@noop {}
	{\bibfield  {journal} {\bibinfo  {journal} {Adv. Funct. Mater.}\ ,\ \bibinfo
			{pages} {2505145}} (\bibinfo {year} {2025})}\BibitemShut {NoStop}%
	\bibitem [{\citenamefont {Wang}\ \emph {et~al.}()\citenamefont {Wang},
		\citenamefont {Li},\ and\ \citenamefont {Yang}}]{wang2025two}%
	\BibitemOpen
	\bibfield  {author} {\bibinfo {author} {\bibfnamefont {Y.-K.}\ \bibnamefont
			{Wang}}, \bibinfo {author} {\bibfnamefont {S.}~\bibnamefont {Li}},\ and\
		\bibinfo {author} {\bibfnamefont {S.~A.}\ \bibnamefont {Yang}},\ }\bibfield
	{title} {\bibinfo {title} {Two-dimensional altermagnetic iron oxyhalides:
			Real chern topology and valley-spin-lattice coupling},\ }\href@noop {}
	{\bibinfo  {journal} {arXiv:2510.12748}\ }\BibitemShut {NoStop}%
	\bibitem [{\citenamefont {Kresse}\ and\ \citenamefont
		{Hafner}(1994)}]{kresse1994}%
	\BibitemOpen
	\bibfield  {journal} {  }\bibfield  {author} {\bibinfo {author} {\bibfnamefont
			{G.}~\bibnamefont {Kresse}}\ and\ \bibinfo {author} {\bibfnamefont
			{J.}~\bibnamefont {Hafner}},\ }\bibfield  {title} {\bibinfo {title} {Ab
			initio molecular-dynamics simulation of the
			liquid-metal–amorphous-semiconductor transition in germanium},\ }\href@noop
	{} {\bibfield  {journal} {\bibinfo  {journal} {Phys. Rev. B}\ }\textbf
		{\bibinfo {volume} {49}},\ \bibinfo {pages} {14251} (\bibinfo {year}
		{1994})}\BibitemShut {NoStop}%
	\bibitem [{\citenamefont {Kresse}\ and\ \citenamefont
		{Furthm{\"u}ller}(1996)}]{kresse1996}%
	\BibitemOpen
	\bibfield  {author} {\bibinfo {author} {\bibfnamefont {G.}~\bibnamefont
			{Kresse}}\ and\ \bibinfo {author} {\bibfnamefont {J.}~\bibnamefont
			{Furthm{\"u}ller}},\ }\bibfield  {title} {\bibinfo {title} {Efficient
			iterative schemes for ab initio total-energy calculations using a plane-wave
			basis set},\ }\href@noop {} {\bibfield  {journal} {\bibinfo  {journal} {Phys.
				Rev. B}\ }\textbf {\bibinfo {volume} {54}},\ \bibinfo {pages} {11169}
		(\bibinfo {year} {1996})}\BibitemShut {NoStop}%
	\bibitem [{\citenamefont {Bl{\"o}chl}(1994)}]{blochl1994projector}%
	\BibitemOpen
	\bibfield  {author} {\bibinfo {author} {\bibfnamefont {P.~E.}\ \bibnamefont
			{Bl{\"o}chl}},\ }\bibfield  {title} {\bibinfo {title} {Projector
			augmented-wave method},\ }\href@noop {} {\bibfield  {journal} {\bibinfo
			{journal} {Phys. Rev. B}\ }\textbf {\bibinfo {volume} {50}},\ \bibinfo
		{pages} {17953} (\bibinfo {year} {1994})}\BibitemShut {NoStop}%
	\bibitem [{\citenamefont {Perdew}\ \emph {et~al.}(1996)\citenamefont {Perdew},
		\citenamefont {Burke},\ and\ \citenamefont {Ernzerhof}}]{PBE}%
	\BibitemOpen
	\bibfield  {author} {\bibinfo {author} {\bibfnamefont {J.~P.}\ \bibnamefont
			{Perdew}}, \bibinfo {author} {\bibfnamefont {K.}~\bibnamefont {Burke}},\ and\
		\bibinfo {author} {\bibfnamefont {M.}~\bibnamefont {Ernzerhof}},\ }\bibfield
	{title} {\bibinfo {title} {Generalized gradient approximation made simple},\
	}\href@noop {} {\bibfield  {journal} {\bibinfo  {journal} {Phys. Rev. Lett.}\
		}\textbf {\bibinfo {volume} {77}},\ \bibinfo {pages} {3865} (\bibinfo {year}
		{1996})}\BibitemShut {NoStop}%
	\bibitem [{\citenamefont {Anisimov}\ \emph {et~al.}(1991)\citenamefont
		{Anisimov}, \citenamefont {Zaanen},\ and\ \citenamefont
		{Andersen}}]{Anisimov1991}%
	\BibitemOpen
	\bibfield  {author} {\bibinfo {author} {\bibfnamefont {V.~I.}\ \bibnamefont
			{Anisimov}}, \bibinfo {author} {\bibfnamefont {J.}~\bibnamefont {Zaanen}},\
		and\ \bibinfo {author} {\bibfnamefont {O.~K.}\ \bibnamefont {Andersen}},\
	}\bibfield  {title} {\bibinfo {title} {Band theory and mott insulators:
			Hubbard $\mathrm{U}$ instead of stoner $\mathrm{I}$},\ }\href@noop {}
	{\bibfield  {journal} {\bibinfo  {journal} {Phys. Rev. B}\ }\textbf {\bibinfo
			{volume} {44}},\ \bibinfo {pages} {943} (\bibinfo {year} {1991})}\BibitemShut
	{NoStop}%
	\bibitem [{\citenamefont {Dudarev}\ \emph {et~al.}(1998)\citenamefont
		{Dudarev}, \citenamefont {Botton}, \citenamefont {Savrasov}, \citenamefont
		{Humphreys},\ and\ \citenamefont {Sutton}}]{dudarev1998}%
	\BibitemOpen
	\bibfield  {author} {\bibinfo {author} {\bibfnamefont {S.~L.}\ \bibnamefont
			{Dudarev}}, \bibinfo {author} {\bibfnamefont {G.~A.}\ \bibnamefont {Botton}},
		\bibinfo {author} {\bibfnamefont {S.~Y.}\ \bibnamefont {Savrasov}}, \bibinfo
		{author} {\bibfnamefont {C.~J.}\ \bibnamefont {Humphreys}},\ and\ \bibinfo
		{author} {\bibfnamefont {A.~P.}\ \bibnamefont {Sutton}},\ }\bibfield  {title}
	{\bibinfo {title} {Electron-energy-loss spectra and the structural stability
			of nickel oxide: An $\mathrm{LSDA+U}$ study},\ }\href@noop {} {\bibfield
		{journal} {\bibinfo  {journal} {Phys. Rev. B}\ }\textbf {\bibinfo {volume}
			{57}},\ \bibinfo {pages} {1505} (\bibinfo {year} {1998})}\BibitemShut
	{NoStop}%
	\bibitem [{\citenamefont {Jain}\ \emph {et~al.}(2011)\citenamefont {Jain},
		\citenamefont {Hautier}, \citenamefont {Ong}, \citenamefont {Moore},
		\citenamefont {Fischer}, \citenamefont {Persson},\ and\ \citenamefont
		{Ceder}}]{jain2011formation}%
	\BibitemOpen
	\bibfield  {author} {\bibinfo {author} {\bibfnamefont {A.}~\bibnamefont
			{Jain}}, \bibinfo {author} {\bibfnamefont {G.}~\bibnamefont {Hautier}},
		\bibinfo {author} {\bibfnamefont {S.~P.}\ \bibnamefont {Ong}}, \bibinfo
		{author} {\bibfnamefont {C.~J.}\ \bibnamefont {Moore}}, \bibinfo {author}
		{\bibfnamefont {C.~C.}\ \bibnamefont {Fischer}}, \bibinfo {author}
		{\bibfnamefont {K.~A.}\ \bibnamefont {Persson}},\ and\ \bibinfo {author}
		{\bibfnamefont {G.}~\bibnamefont {Ceder}},\ }\bibfield  {title} {\bibinfo
		{title} {Formation enthalpies by mixing $\mathrm{GGA}$ and $\mathrm{GGA+U}$
			calculations},\ }\href@noop {} {\bibfield  {journal} {\bibinfo  {journal}
			{Phys. Rev. B}\ }\textbf {\bibinfo {volume} {84}},\ \bibinfo {pages} {045115}
		(\bibinfo {year} {2011})}\BibitemShut {NoStop}%
	\bibitem [{SM()}]{SM}%
	\BibitemOpen
	\href@noop {} {\bibinfo  {journal} {See Supplemental Material for details on
			the magnetic configurations, magnetic transition temperatures, and band
			structure results under different Hubbard $U$ values}\ }\BibitemShut
	{NoStop}%
	\bibitem [{\citenamefont {Togo}\ and\ \citenamefont
		{Tanaka}(2015)}]{togo2015first}%
	\BibitemOpen
	\bibfield  {journal} {  }\bibfield  {author} {\bibinfo {author} {\bibfnamefont
			{A.}~\bibnamefont {Togo}}\ and\ \bibinfo {author} {\bibfnamefont
			{I.}~\bibnamefont {Tanaka}},\ }\bibfield  {title} {\bibinfo {title} {First
			principles phonon calculations in materials science},\ }\href@noop {}
	{\bibfield  {journal} {\bibinfo  {journal} {Scr. Mater.}\ }\textbf {\bibinfo
			{volume} {108}},\ \bibinfo {pages} {1} (\bibinfo {year} {2015})}\BibitemShut
	{NoStop}%
	\bibitem [{\citenamefont {Pizzi}\ \emph {et~al.}(2020)\citenamefont {Pizzi},
		\citenamefont {Vitale}, \citenamefont {Arita}, \citenamefont {Bl{\"u}gel},
		\citenamefont {Freimuth}, \citenamefont {G{\'e}ranton}, \citenamefont
		{Gibertini}, \citenamefont {Gresch}, \citenamefont {Johnson}, \citenamefont
		{Koretsune} \emph {et~al.}}]{pizzi2020wannier90}%
	\BibitemOpen
	\bibfield  {author} {\bibinfo {author} {\bibfnamefont {G.}~\bibnamefont
			{Pizzi}}, \bibinfo {author} {\bibfnamefont {V.}~\bibnamefont {Vitale}},
		\bibinfo {author} {\bibfnamefont {R.}~\bibnamefont {Arita}}, \bibinfo
		{author} {\bibfnamefont {S.}~\bibnamefont {Bl{\"u}gel}}, \bibinfo {author}
		{\bibfnamefont {F.}~\bibnamefont {Freimuth}}, \bibinfo {author}
		{\bibfnamefont {G.}~\bibnamefont {G{\'e}ranton}}, \bibinfo {author}
		{\bibfnamefont {M.}~\bibnamefont {Gibertini}}, \bibinfo {author}
		{\bibfnamefont {D.}~\bibnamefont {Gresch}}, \bibinfo {author} {\bibfnamefont
			{C.}~\bibnamefont {Johnson}}, \bibinfo {author} {\bibfnamefont
			{T.}~\bibnamefont {Koretsune}}, \emph {et~al.},\ }\bibfield  {title}
	{\bibinfo {title} {Wannier90 as a community code: new features and
			applications},\ }\href@noop {} {\bibfield  {journal} {\bibinfo  {journal} {J.
				Phys.: Condens. Matter}\ }\textbf {\bibinfo {volume} {32}},\ \bibinfo {pages}
		{165902} (\bibinfo {year} {2020})}\BibitemShut {NoStop}%
	\bibitem [{\citenamefont {Lin}\ \emph {et~al.}(2018)\citenamefont {Lin},
		\citenamefont {Si}, \citenamefont {Zhu}, \citenamefont {Cai}, \citenamefont
		{Li}, \citenamefont {Kong}, \citenamefont {Yu},\ and\ \citenamefont
		{Wen}}]{lin2018structure}%
	\BibitemOpen
	\bibfield  {author} {\bibinfo {author} {\bibfnamefont {H.}~\bibnamefont
			{Lin}}, \bibinfo {author} {\bibfnamefont {J.}~\bibnamefont {Si}}, \bibinfo
		{author} {\bibfnamefont {X.}~\bibnamefont {Zhu}}, \bibinfo {author}
		{\bibfnamefont {K.}~\bibnamefont {Cai}}, \bibinfo {author} {\bibfnamefont
			{H.}~\bibnamefont {Li}}, \bibinfo {author} {\bibfnamefont {L.}~\bibnamefont
			{Kong}}, \bibinfo {author} {\bibfnamefont {X.}~\bibnamefont {Yu}},\ and\
		\bibinfo {author} {\bibfnamefont {H.-H.}\ \bibnamefont {Wen}},\ }\bibfield
	{title} {\bibinfo {title} {Structure and physical properties of
			$\mathrm{CsV_{2}Se_{2-x}O}$ and $\mathrm{V_{2}Se_{2}O}$},\ }\href@noop {}
	{\bibfield  {journal} {\bibinfo  {journal} {Phys. Rev. B}\ }\textbf {\bibinfo
			{volume} {98}},\ \bibinfo {pages} {075132} (\bibinfo {year}
		{2018})}\BibitemShut {NoStop}%
	\bibitem [{\citenamefont {Ablimit}\ \emph {et~al.}(2018)\citenamefont
		{Ablimit}, \citenamefont {Sun}, \citenamefont {Cheng}, \citenamefont {Liu},
		\citenamefont {Wu}, \citenamefont {Jiang}, \citenamefont {Ren}, \citenamefont
		{Li},\ and\ \citenamefont {Cao}}]{ablimit2018v2te2o}%
	\BibitemOpen
	\bibfield  {author} {\bibinfo {author} {\bibfnamefont {A.}~\bibnamefont
			{Ablimit}}, \bibinfo {author} {\bibfnamefont {Y.-L.}\ \bibnamefont {Sun}},
		\bibinfo {author} {\bibfnamefont {E.-J.}\ \bibnamefont {Cheng}}, \bibinfo
		{author} {\bibfnamefont {Y.-B.}\ \bibnamefont {Liu}}, \bibinfo {author}
		{\bibfnamefont {S.-Q.}\ \bibnamefont {Wu}}, \bibinfo {author} {\bibfnamefont
			{H.}~\bibnamefont {Jiang}}, \bibinfo {author} {\bibfnamefont
			{Z.}~\bibnamefont {Ren}}, \bibinfo {author} {\bibfnamefont {S.}~\bibnamefont
			{Li}},\ and\ \bibinfo {author} {\bibfnamefont {G.-H.}\ \bibnamefont {Cao}},\
	}\bibfield  {title} {\bibinfo {title} {$\mathrm{V_{2}Te_{2}O}$: a
			two-dimensional van der waals correlated metal},\ }\href@noop {} {\bibfield
		{journal} {\bibinfo  {journal} {Inorg. Chem.}\ }\textbf {\bibinfo {volume}
			{57}},\ \bibinfo {pages} {14617} (\bibinfo {year} {2018})}\BibitemShut
	{NoStop}%
	\bibitem [{\citenamefont {Ahmed}\ \emph {et~al.}(2015)\citenamefont {Ahmed},
		\citenamefont {Modine},\ and\ \citenamefont {Zhu}}]{ahmed2015bonding}%
	\BibitemOpen
	\bibfield  {author} {\bibinfo {author} {\bibfnamefont {T.}~\bibnamefont
			{Ahmed}}, \bibinfo {author} {\bibfnamefont {N.}~\bibnamefont {Modine}},\ and\
		\bibinfo {author} {\bibfnamefont {J.-X.}\ \bibnamefont {Zhu}},\ }\bibfield
	{title} {\bibinfo {title} {Bonding between graphene and $\mathrm{MoS_{2}}$
			monolayers without and with li intercalation},\ }\href@noop {} {\bibfield
		{journal} {\bibinfo  {journal} {Appl. Phys. Lett.}\ }\textbf {\bibinfo
			{volume} {107}},\ \bibinfo {pages} {043903} (\bibinfo {year}
		{2015})}\BibitemShut {NoStop}%
	\bibitem [{\citenamefont {Lahourpour}\ \emph {et~al.}(2019)\citenamefont
		{Lahourpour}, \citenamefont {Boochani}, \citenamefont {Parhizgar},\ and\
		\citenamefont {Elahi}}]{lahourpour2019structural}%
	\BibitemOpen
	\bibfield  {author} {\bibinfo {author} {\bibfnamefont {F.}~\bibnamefont
			{Lahourpour}}, \bibinfo {author} {\bibfnamefont {A.}~\bibnamefont
			{Boochani}}, \bibinfo {author} {\bibfnamefont {S.}~\bibnamefont
			{Parhizgar}},\ and\ \bibinfo {author} {\bibfnamefont {S.}~\bibnamefont
			{Elahi}},\ }\bibfield  {title} {\bibinfo {title} {Structural, electronic and
			optical properties of graphene-like nano-layers $\mathrm{MoS_{2}}
			(\mathrm{X}: \mathrm{S, Se, Te}): \mathrm{DFT}$ study},\ }\href@noop {}
	{\bibfield  {journal} {\bibinfo  {journal} {J. Theor. Appl. Phys.}\ }\textbf
		{\bibinfo {volume} {13}},\ \bibinfo {pages} {191} (\bibinfo {year}
		{2019})}\BibitemShut {NoStop}%
	\bibitem [{\citenamefont {Ermolaev}\ \emph {et~al.}(2020)\citenamefont
		{Ermolaev}, \citenamefont {Stebunov}, \citenamefont {Vyshnevyy},
		\citenamefont {Tatarkin}, \citenamefont {Yakubovsky}, \citenamefont
		{Novikov}, \citenamefont {Baranov}, \citenamefont {Shegai}, \citenamefont
		{Nikitin}, \citenamefont {Arsenin} \emph {et~al.}}]{ermolaev2020broadband}%
	\BibitemOpen
	\bibfield  {author} {\bibinfo {author} {\bibfnamefont {G.~A.}\ \bibnamefont
			{Ermolaev}}, \bibinfo {author} {\bibfnamefont {Y.~V.}\ \bibnamefont
			{Stebunov}}, \bibinfo {author} {\bibfnamefont {A.~A.}\ \bibnamefont
			{Vyshnevyy}}, \bibinfo {author} {\bibfnamefont {D.~E.}\ \bibnamefont
			{Tatarkin}}, \bibinfo {author} {\bibfnamefont {D.~I.}\ \bibnamefont
			{Yakubovsky}}, \bibinfo {author} {\bibfnamefont {S.~M.}\ \bibnamefont
			{Novikov}}, \bibinfo {author} {\bibfnamefont {D.~G.}\ \bibnamefont
			{Baranov}}, \bibinfo {author} {\bibfnamefont {T.}~\bibnamefont {Shegai}},
		\bibinfo {author} {\bibfnamefont {A.~Y.}\ \bibnamefont {Nikitin}}, \bibinfo
		{author} {\bibfnamefont {A.~V.}\ \bibnamefont {Arsenin}}, \emph {et~al.},\
	}\bibfield  {title} {\bibinfo {title} {Broadband optical properties of
			monolayer and bulk $\mathrm{MoS_{2}}$},\ }\href@noop {} {\bibfield  {journal}
		{\bibinfo  {journal} {npj 2D Mater. Appl.}\ }\textbf {\bibinfo {volume}
			{4}},\ \bibinfo {pages} {1} (\bibinfo {year} {2020})}\BibitemShut {NoStop}%
	\bibitem [{\citenamefont {Li}\ \emph {et~al.}(2020)\citenamefont {Li},
		\citenamefont {Wu}, \citenamefont {Feng}, \citenamefont {Guan}, \citenamefont
		{Feng}, \citenamefont {Yao},\ and\ \citenamefont {Yang}}]{li2020valley}%
	\BibitemOpen
	\bibfield  {author} {\bibinfo {author} {\bibfnamefont {S.}~\bibnamefont
			{Li}}, \bibinfo {author} {\bibfnamefont {W.}~\bibnamefont {Wu}}, \bibinfo
		{author} {\bibfnamefont {X.}~\bibnamefont {Feng}}, \bibinfo {author}
		{\bibfnamefont {S.}~\bibnamefont {Guan}}, \bibinfo {author} {\bibfnamefont
			{W.}~\bibnamefont {Feng}}, \bibinfo {author} {\bibfnamefont {Y.}~\bibnamefont
			{Yao}},\ and\ \bibinfo {author} {\bibfnamefont {S.~A.}\ \bibnamefont
			{Yang}},\ }\bibfield  {title} {\bibinfo {title} {Valley-dependent properties
			of monolayer $\mathrm{MoSi_{2}N_{4}}$, $\mathrm{WSi_{2}N_{4}}$, and
			$\mathrm{MoSi_{2}As_{4}}$},\ }\href@noop {} {\bibfield  {journal} {\bibinfo
			{journal} {Phys. Rev. B}\ }\textbf {\bibinfo {volume} {102}},\ \bibinfo
		{pages} {235435} (\bibinfo {year} {2020})}\BibitemShut {NoStop}%
\end{thebibliography}

%

\end{document}